\documentclass[12pt]{article}
\usepackage{epsfig}
\def\be{\begin{equation}}
\def\ee{\end{equation}}
\def\bq{\begin{eqnarray}}
\def\eq{\end{eqnarray}}

\newcommand{\sle}{e \kern-6pt/}
\newcommand{\slk}{/\kern-6pt k}
\newcommand{\sll}{/\kern-4pt l}
\newcommand{\slp}{p\kern-5pt/}
\newcommand{\slq}{q\kern-5.5pt/}
\newcommand{\slep}{\epsilon \kern-6pt/}

\begin{document}
\thispagestyle{empty}
\begin{flushright}
MCTP-02-66\\
%(Version of~\today)\\
\end{flushright}
\vspace{0.5cm}

\begin{center} 
{\Large \bf 
On the Resummation of Large QCD Corrections to $\gamma \gamma \to b \bar b$
}

\vspace{1.7cm}
{\sc \bf R. Akhoury$^{a}$, H. Wang$^{b}$ and O. Yakovlev$^{c}$}\\[1cm]
\end{center}
\begin{center} \em 
Michigan Center for Theoretical Physics\\
Randall Laboratory of Physics, University of Michigan, \\
Ann Arbor,  Michigan 48109-1120, USA
\end{center}
\vspace{2cm}
% abstract ---------------------------------------
\begin{abstract}\noindent
{ 
 We study the resummation of large QCD radiative corrections up to the next to
leading logarithmic accuracy to the process  $\gamma \gamma \to b \bar b$; i.e.,
we resum logarithms of the type $\alpha_s^p\ln^{2p}{m^2/s}$ and
$\alpha_s^p\ln^{2p-1}{m^2/s}$ ($m$ is the quark mass). The only source of all
the logarithms to this accuracy is the off-shell Sudakov form 
factor included into the triangle topologies of the 
one-loop box diagram. We prove that any other configurations of diagrams  
to this accuracy, either cancel in subgroups or develop a universal on-shell
Sudakov exponent due to the final quark anti-quark lines. 
We study the mechanism of cancellations between the different diagrams, 
which leads to the simple resummed  results.  
We show the cancellation explicitly at three loops for the leading and
at two loops for the next-to-leading logarithms. We also
point out the general mechanism responsible for it, and discuss how it can
be extended to higher orders.}
\end{abstract}
\vspace{4cm}
%\vspace*{\fill}
%\hline\\
% footnotes -------------------------------------
\vspace{1cm}
\noindent $^a${\small e-mail: akhoury@umich.edu}\\
\noindent $^b${\small e-mail: haibinw@umich.edu}\\
\noindent $^c${\small e-mail: yakovlev@umich.edu}
% main text ------------------------------------
\newpage
\section{Introduction}
Future Linear Colliders are expected to reveal the answers to many questions of
modern particle physics. One of these concerns the the physics of the Higgs
particle  and origin of electroweak symmetry breaking. 
The neutral scalar Higgs boson is an important ingredient of the
Standard Model (SM) and is the only SM elementary particle
which has not been detected so far  
(see for a review \cite{Gunion:1989we}, \cite{Gunion:1992ce}).  
The lower limit on $m_H$ of approximately $113.5$ GeV at $95\%$  c.l. 
has been obtained from direct searches at LEP \cite{Hagiwara:fs}. 
Current experiments are concentrating on the possibility of finding a
Higgs particle in the intermediate mass region 
 $113.5 < m_H < 150\quad\mbox{GeV}.$ 
In this region it decays mainly to a $b\bar b$ pair. 
  
The {\it photon  mode} of the future Linear Collider (LC), namely the 
collisions of the energetic polarized Compton photons, 
will be used for the production and for the study of the Higgs particle. 
In the intermediate mass range, the main production process is 
 $$\gamma\gamma\to H \to b\bar b.$$ 
QCD as well as electroweak radiative corrections to this process 
have been studied very well and have been found to be small in this region 
\cite{QCDEW}.    
The main challenge, however, is to get under control the background
process $$\gamma \gamma \to b \bar b,$$ which gets extremely large QCD 
corrections.

 In this paper we discuss the process of the quark anti-quark
production in the  photon mode of the LC, $\gamma \gamma \to b \bar b$. 
The amplitude for this process in the scalar channel contains 
large double logarithms at $|s|,|t|,|u| \gg m^2$.  
At very high energies, the large logarithms spoil the perturbative  
predictions. Therefore it is mandatory to develop a clear resummation
procedure for this double and single logarithmic terms. 
Let us stress, that the main interest in the process
$\gamma \gamma \to b \bar b$ comes from the fact (but is not limited to it) 
that it represents the dominant background for the production  
of the Higgs particle,   
$\gamma \gamma \to H \to b \bar b$
\cite{Ginzburg:1981vm,Telnov:hc,Fadin:1997sn}.
In fact, our motivations for this study are twofold:\\ 
(1) A detailed study of the process $\gamma \gamma \to b \bar b$  
is very important due to  phenomenological  reasons mentioned above; and\\
(2) in addition to that, the quark anti-quark production 
in photon collisions , being one of the simplest process in QCD,
is important in its own right, e.g. for studying and understanding 
QCD effects. 

The Born cross section for the polarized 
$ b\bar b$  production in the scalar channel (at $J_z=0$), 
where the Higgs will be studied as well, is suppressed 
by $\frac{m_b^2}{s}$  \cite{Borden:1994fv, Jikia:1996bi}, 
(here $s$ is the center of mass energy of the initial photons). 
However, the perturbative QCD corrections contains the large double
logarithms of the form  $\rho=\alpha_s\ln^2 (\frac{m_b^2}{s})$, which 
give a contribution to the cross section which is of the same order as the Born 
contribution at high energies. 
 The presence of the large correction was noticed by Jikia 
in \cite{Jikia:1996bi}. 
The double logarithmic nature and the  origin of these corrections 
were studied in \cite{Fadin:1997sn}.  
The authors studied the process to one and two loop accuracy.    
Later, the form of the resummed results for the double logarithms has been 
argued in \cite{Melles:1998gu}. These authors also claimed that the double logarithms 
have a``non-Sudakov'' origin. 
 
In this paper, we first present an alternative way of understanding 
the resummation procedure for the double logarithms. The general idea of our
approach, is that the only source of double logarithms is 
the off-shell Sudakov form factor included in the triangle topologies of the
one-loop box diagram. We have proved that the other types of the higher loop
diagrams will either cancel in subgroups or develop a universal on-shell
Sudakov exponent due to the final quark anti-quark lines. 
In addition to that, (1) we extend this analysis to the
next-to-leading-logarithmic (NLL) accuracy, and (2)  
we study the mechanism of cancellations alluded to above,
between the different diagrams which leads to very simple resummed
results.  We demonstrate the cancellation explicitly at three loops and
point out the mechanism responsible which then allows for the generalization 
to higher loops, all up to the next to leading logarithmic accuracy.  
As an aside, we argue that all the large logarithms up to the next to 
leading level
are related to the Sudakov ones. This includes not only the leading ones of the
form $\alpha_s^p\ln^{2p}{m^2/s}$ but also the next to leading ones of the form
$\alpha_s^p\ln^{2p-1}{m^2/s}$ ($m$ is the quark mass). It is this fact together
with an understanding of the cancellation mechanism which allows us to
develop an easy resummation procedure. 

The paper is organized as follows. In section 2.1 we
discuss the one loop diagrams and explain which topologies are important. In
section 2.2 we present a procedure for the resummation of the double logarithms.
In section 2.3 we extend our analysis to the next-to-leading logarithmic level. 
In this section we consider all possible topologies and give the final
result of the resummation. Section 3 is devoted to the study of the different 
cancellations which are responsible for the simple resummed results of section
2.2 for the double logarithmic case. In section 4 we tackle the case of the form
factors to the next to leading logarithmic accuracy and justify the results given
in section 2.3. The paper ends with some conclusions and discussions.

\section{The resummation up to next to leading logarithms}

\subsection{One loop results}
We study the process of quark anti-quark pair production in photon-photon collisions  
\begin{eqnarray}
\gamma (k_1) + \gamma (k_2) \to q (p_1)+\bar{q}(p_2)
\end{eqnarray}
in the scalar channel $J_z=0$, 
at very high energies compared to the mass of the quark $m$, 
and at large scattering angles 
\begin{equation}
|s| \sim |t| \sim |u|\gg m^2.
\end{equation}
The radiative corrections to this helicity amplitude contain  
 large double and single QCD logarithms of the form 
$\alpha_s \ln^p (s/m^2), p=2,1$, in the limit
 eq.(\ref{lim1},\ref{lim2}).

\begin{figure}
\centerline{\epsfig{file=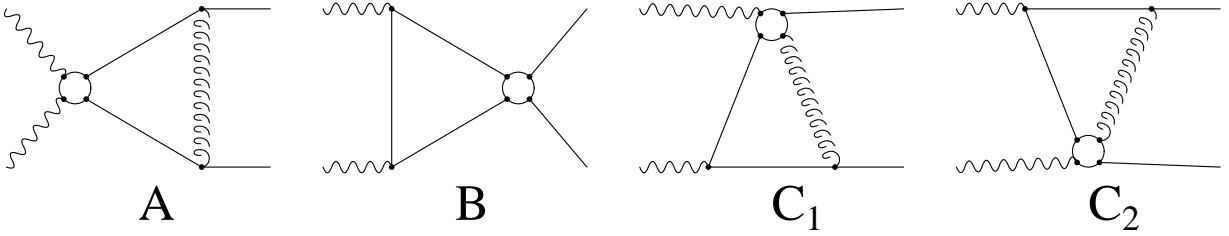,scale=0.85}}
\caption{\label{Topologies}
Triangle topologies obtained from the one loop box diagram.}
\end{figure}

It has been shown in \cite{Fadin:1997sn} that the large double logarithms (DL) have 
a Sudakov-like nature and can be extracted from the total 
result by identifying special kinematic regions. It was demonstrated that,  
only the box diagram contributes, and that this box diagram can be reduced 
to three different effective diagrams with triangle 
topology.  We label them as topologies A,B,C following \cite{Fadin:1997sn}, see Fig.(\ref{Topologies}). 
These effective diagrams result from the box diagram 
when one of the four propagators is hard, 
two collinear and one soft ( the soft line can be either the gluon 
or the quark propagator). Hence one can 
shrink the hard propagator of the box diagram to a point, because it does not
depend on the soft loop momenta anymore. The hard momenta would be just
an ultra violet cutoff in this situation.  

In order to introduce notations and to get an idea about the structure of the
DL, let us consider one loop calculations in the DL approximation. 
Also, we will use these results in the next section for the 
construction of the DL-resummed result.  
 
As is well known, the double logarithms come from the region of soft 
loop momenta. Hence, one can always write 
a factorized form for the amplitude in the DL approximation:  
\begin{eqnarray}
M_A=M_{\mbox{Born}}\cdot F_A,\qquad M_B=M_{\mbox{Born}}\cdot F_B,\qquad
 M_C=2 M_{\mbox{Born}}\cdot F_C,
\end{eqnarray}
the factor 2 in the last equation represents two identical contributions
$C_1, C_2$ of the topology C. 
The amplitude then reads
\begin{eqnarray}
M=M_{\mbox{Born}}\cdot \Big( F_A+F_B+2 F_C \Big).
\end{eqnarray}
We turn now to the calculation of the form factors $F_A,F_B,F_C$. 

It is convenient to use the standard Sudakov technique 
for calculating DL contributions, as described in 
\cite{Sudakov:1954sw,Fadin:1997sn},  
and introduce Sudakov parameterization \cite{Sudakov:1954sw,Gorshkov:ht}. 
One starts with decomposing the soft momenta in terms of those along the hard 
external momenta, for example $k_1, k_2$  and transverse to them
\begin{eqnarray}
k=\alpha k_1 +\beta k_2 + k_{\bot}.
\end{eqnarray}
The DL contribution comes only from the region  
\begin{eqnarray}\label{kinem1}
m^2, |k_{\bot}|^2 \ll s|\alpha|,s|\beta| \ll s.
\end{eqnarray}
For different topologies it is convenient to use different
decompositions of the soft momenta:
\begin{eqnarray}
\mbox{topology A:} \qquad k&=& \beta (p- \frac{m^2}{s}\bar p)+
\alpha (\bar p - \frac{m^2}{s} p ) +k_{\bot}\\
\mbox{topology B:}\qquad k&=& \beta k_2 +
\alpha k_1 + k_{\bot}\\ 
\mbox{topology C:}\qquad k&=&\beta k_2 +
\alpha (\bar p - \frac{m^2}{s} k_2)  + k_{\bot} 
\end{eqnarray} 
The loop integration in terms of the new variables reads 
\begin{eqnarray}
\int\limits_{-\infty}^{\infty} d^4k
=\frac{s}{2}\int\limits_{-\infty}^{\infty}d\alpha
\int\limits_{-\infty}^{\infty} d\beta \int\limits_{0}^{\infty}\pi dl^2_{\bot}.
\end{eqnarray}
The integration over the transverse momenta of the soft quark is performed
by taking half of the residues in the corresponding propagator, 
\begin{eqnarray}
\int \frac{d^4k}{k^2-\lambda^2+i0} F=\int\frac{s}{2}
\frac{d\alpha d\beta d^2k_\bot}{s\alpha\beta-k_\bot^2-\lambda^2+i0} F \to
-i\pi^2\frac{s}{2}\int d\alpha d\beta \Theta (s\alpha\beta-\lambda^2)F,
\end{eqnarray}
here $\lambda$ is a small (fictitious) gluon mass. 
In this manner the one loop amplitude may be calculated 
in the DL approximation and topology A gives,
\begin{eqnarray}\label{1loopA}
F^{1-loop}_A&=&-\frac{C_F\alpha_s}{2\pi}
\int\limits_{0}^{1} \int\limits_{0}^{1}\frac{d\alpha d\beta }{\alpha \beta} 
\Theta(\alpha \beta -\frac{\lambda^2}{s} )
\Theta(\alpha -\frac{\lambda^2}{s}\beta )
\Theta(\beta -\frac{\lambda^2}{s}\alpha )\\
&=&-\frac{C_F\alpha_s}{2\pi}\Big( \frac12\ln^2 \frac{m^2}{s} +
\ln \frac{m^2}{s}\ln \frac{\lambda^2}{m^2} \Big),
\end{eqnarray}
Topology B and C have following form at one loop in the DL approximation
\begin{eqnarray}\label{1loopB}
F^{1-loop}_B=F^{1-loop}_C&=&-\frac{C_F\alpha_s}{2\pi}
\int\limits_{0}^{1}\int\limits_{0}^{1} 
\frac{d\alpha d\beta }{\alpha \beta} 
\Theta(\alpha \beta -\frac{m^2}{s} )
=-\frac{C_F\alpha_s}{2\pi}\Big( \frac12\ln^2 \frac{m^2}{s}\Big)
\end{eqnarray}
These results  were derived first in \cite{Fadin:1997sn}  
and will be used in our calculation later on.

\subsection{The double logarithmic resummation}

The main idea behind our method is that
the only origin of the double logarithms is the off-shell Sudakov 
form factor \footnote{ hard outer lines are slightly off-shell} 
included in the effective one loop triangle diagrams A,B and C. 
Contributions from other types of diagrams will either cancel in subgroups  
or develop a simple on-shell Sudakov exponent due to the final quark 
anti-quark pair.
The cancellations amongst the diagrams will be discussed in the next
section where the general mechanism is elucidated. Here we study the DL resummation 
in the topologies A, B and C.     

The diagrams of topology A represent a simple case. 
Here we do not have  any ``hard''
 DLs, but only the ``soft'' ones, which develop the usual on-shell 
Sudakov form factor from the gluon exchanges between the final quark antiquark pair.
It is well known, that the on-shell DL Sudakov form factor in QCD is  
the exponent of the one loop result  
\begin{eqnarray}\label{topA}
F_A = \mbox{Exp}\Big( F^{1-loop}_A \Big),
\end{eqnarray}
with $F^{1-loop}_A$ given in eq.(\ref{1loopA}). 

The resummation of the DL terms coming from the diagrams of 
topology B is more involved.  
First, we have to account for the ``soft'' on-shell exponent from 
the final quark antiquark re-scattering, similar to the one in topology A, 
eq.(\ref{topA}). 
Second, we observe a new element, namely the ``hard'' double logarithms 
from quark anti-quark re-scattering inside the one loop diagram. 
The interacting particles (quarks in this case) are hard and
slightly off-shell, $p_{1,2}^2\ll s$. 
  
 According to our general idea, the resummation of the ``hard'' DL   
can be obtained by including the off-shell Sudakov form factor 
into the loop, and accounting only for those momenta which do not 
destroy the DL nature of the one loop result. 
 
We recall that the off-shell Sudakov form factor, is a vertex of 
the production of the quark and anti-quark with small virtualities 
$p_1^2$ and $p_2^2$ and has a form \cite{Carazzone:hj} 
%\cite{Sudakov:1954sw,Smilga:uj} 
\begin{eqnarray}\label{SudB}
S(p_1 ,p_2)=\mbox{Exp}\Big( -\frac{C_F\alpha_s}{2\pi}
\ln(\frac{s}{ |p_1|^2})\ln(\frac{s}{ |p_2|^2}) \Big),
\end{eqnarray}
assuming  that 
%\begin{eqnarray}
$m^2 \ll |p_1|^2, |p_2|^2 \ll s.$
%\end{eqnarray}
Now the momenta $p_1$ and $p_2$ become loop momenta, i.e.,they depend on
$\alpha, \beta$
\begin{eqnarray}\label{lim1}
p_1^2=(k_1+k)^2=s\beta,\qquad p_2^2=(k_2-k)=-s\alpha.
\end{eqnarray}
Because of the DL approximation, the kinematic region 
of interest is restricted by the inequalities 
\begin{eqnarray}\label{lim2}
m^2 \ll |p_1|^2, |p_2|^2 \ll s.
\end{eqnarray}
Taking into account the result eq.(\ref{SudB}) and including it into eq.(\ref{1loopB}) we get 
\begin{eqnarray}
F_B&=&-\frac{C_F\alpha_s}{2\pi}
\int\limits_{0}^{1}\int\limits_{0}^{1} 
\frac{d\alpha d\beta }{\alpha \beta} 
\Theta(\alpha \beta -\frac{m^2}{s} )
\mbox{Exp}\Big( -\frac{C_F\alpha_s}{2\pi}
\ln |\alpha| \ln |\beta|  \Big)
\end{eqnarray}
We next transform the exponent into a power series 
and find that the integral of the $n-$th term will be of the form
\begin{equation}\label{integral}
\int\limits_0^1 d \xi_1 \int\limits_{1-\xi_1}^1 d \xi_2
\xi_1^{n+a}\xi_2^{n+b}=\frac{\Gamma (n+a+1)\Gamma (n+b+1)}{\Gamma (3+2n+a+b)}.
\end{equation}
The final result at DL accuracy reads
\begin{eqnarray}\label{dlresult}
F_B&=&F_B^{\mbox{1-loop}}\sum\limits_{n=0}^{\infty}
\frac{2\Gamma (n+1)}{\Gamma (2n+3)}
\Big( -\rho_B \Big)^n
\end{eqnarray}
with $\rho_B=\frac{C_F\alpha_s}{2\pi}L^2, L=
\ln\Big(\frac{m^2}{s}\Big)$. 
The index $n$ shows that the order of the amplitude is $\alpha_s^n$. We can 
clearly identify the separate  contributions of the fixed orders in
$\alpha_s$. On the other hand, if $\rho_B $ is large all terms in the
series are important, giving altogether some analytic 
function $F_{DL}(\rho)$. This function is identified with 
a hyper-geometric function $_2F_2(1,1;2,\frac{3}{2};z)$, namely    
\begin{eqnarray}\label{dlr}
F_{B} &=& F^{1-loop}_B \sum\limits_{n=0}^{\infty}
\frac{2\Gamma (n+1)}{\Gamma (2n+3)}
\Big( -\rho_B \Big)^n={}_2F_{2}(1,1;2,\frac{3}{2},-\frac{\rho_B}{4})
F^{1-loop}_B.
\end{eqnarray}
These functions are exact answers in $\rho_B$ in the DL approximation.
For large values of the parameter $\rho$ the function $F_{DL}$ 
has the following asymptotic,  
\begin{eqnarray}
F_B(\rho_B)=\frac{2\ln (2\rho_B )}{\rho_B}F^{1-loop}_B.
\end{eqnarray}
Thus we see that despite the fact that perturbation theory blows up at 
large $\rho_B$, the resummed result gives a smooth well defined function.

Topology C differs from B only through the color structure at the DL level. 
At the end of the calculations the answers for topologies B and C are similar and are 
related by the simple substitution:
\begin{eqnarray}
C_F\to C_A/2\quad \mbox{in the variable}\quad \rho_B.
\end{eqnarray}
Combining  all results for the three topologies we have 
\begin{eqnarray}
M=M_{\mbox{Born}}( 1 + F_B + 2 F_C ) \mbox{Exp}\Big( F^{1-loop}_A \Big).
\end{eqnarray}
with the functions defined above. We must stress that 
there are other diagrams which can give DL contributions. Fortunately, 
all of them cancel in the certain groups, as will be discussed in 
section 3. This cancellation has a very simple physical interpretation 
(see Sec. 3).    

Our results are written in two different representations.
 We have checked that they are in agreement with  
\cite{Melles:1998gu}, where the authors used a completely different
method, not mentioning the off-shell Sudakov form factors at all.    
We believe that our approach is more transparent. It shows that the 
new double logarithms  
are of Sudakov type, and, therefore, the resummation procedure 
becomes simple. Another advantage of our approach is that it 
opens up the possibility for resumming the single logarithms as well. 

\subsection{Next-to-leading-logarithmic accuracy}

It is possible to develop this approach to achieve 
next-to-leading-logarithmic accuracy. 

First, the factorization formula should be modified 
in order to take into account single logarithms. 
The amplitude reads at NLL approximation
\begin{eqnarray}\label{ampNLL}
M=M_{\mbox{Born}}( 1 +\Delta + F_B + 2 F_C ) F_A
\end{eqnarray}
with the function $$\Delta=\frac{\alpha_sC_F}{\pi}\frac{3}{2}\ln
(s/m^2),$$ which can be extracted from the explicit calculations
performed in \cite{Jikia:1996bi} by expanding the result at 
$s=|t|=|u| \gg m^2$. $F_A$ is the on-shell Sudakov form factor to the
next-to-leading logarithmic accuracy \cite{Collins:bt}.

Second, in order to calculate the form factors $F_B$ and $F_C$ at NLL, 
we need an expression for the Sudakov form factor also to this
accuracy.
We turn now to the calculation of the form factors $F_B$ and $F_C$.  
 In fact, such  an analysis already exists in the literature 
\cite{Smilga:uj} for the case when the two external fermion lines are off-shell 
by the same amount, i.e., $p_1^2=p_2^2=p^2$: 
\begin{eqnarray} \label{Smilga}
S_{NLL}(p,p)=\mbox{Exp}\Big( -\frac{C_F\alpha_s(p^2)}{2\pi}
\ln^2(\frac{s}{|p|^2})+\frac{3C_F\alpha_s(p^2)}{4\pi}
\ln (\frac{s}{|p|^2}) \Big).
\end{eqnarray}
For our purposes, we need to
extend the analysis to take into account that $p_1^2 \neq p_2^2$. 
It is easy to see that for the region, eq.(\ref{lim1}), the proof of
factorization and exponentiation given in \cite{Smilga:uj} goes through 
 with straightforward changes. The major change involves the normalization of
the coupling. 
 We have studied one and two loop diagrams for the $q\bar q$ production 
vertex with slightly off-shell quarks with the following result:
\begin{eqnarray}\label{slresult}
S_{NNL}(p_1,p_2)&=&Exp \Big(
-\frac{\alpha_s (\nu^2) C_F}{2\pi} \Big( \ln \frac{|p_1|^2}{s} \ln
\frac{|p_2|^2}{s}
 +\frac{3}{4}\ln \frac{|p_1|^2}{s}
+\frac{3}{4}\ln \frac{|p_2|^2}{s} \Big) \Big), 
\end{eqnarray}  
with the normalization of the coupling constant determined to be  
$\nu^2=\sqrt {|p_1^2| |p^2_2|}$. 
We show only the double and single IR logarithms in the exponent of eq.(\ref{slresult}).

\begin{figure}
\centerline{\epsfig{file=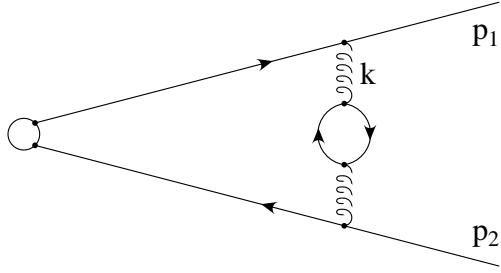, scale=0.85}}
\caption{\label{RGE}
A diagram responsible for determining the normalization scale of the
coupling constant.
}
\end{figure}

In order to understand this normalization,   
we have to consider the diagram shown in Fig.\ref{RGE}

where we keep track of the $n_f$ dependent pieces only since they
are separately gauge invariant.      
Such diagrams can be accounted for by considering the following 
 gluon propagator
\begin{eqnarray}
D^{ab}_{\mu\nu}=-i\delta^{ab}(g_{\mu\nu}-\frac{k_\mu
k_\nu}{k^2})\frac{1}{k^2}
\frac{1}{1+\Pi (k^2)},
\end{eqnarray}
 where $\Pi (k^2)$ is the vacuum polarization by the gluon; 
at the one loop level it is simply $\Pi=\frac{\alpha_s\beta_0}{4\pi}\ln \Big
(\frac{k^2}{\mu^2}e^C\Big), \beta_0=11-\frac{2}{3}n_f$, $C$ being 
a scheme-dependent constant ($\overline{MS}$ scheme $C=-\frac{5}{3}$).  

%The $n_f$ part of this result is gauge invariant
%the essential result may be extracted 
%using gauge invariance and by considering the large $n_f$ limit.
The diagram, Fig.\ref{RGE}, corresponds to the first term in the expansion 
of the gluon propagator in $\alpha_s$. 
The $n_f$--part of this result is as mentioned earlier a 
gauge invariant subset of the complete set of two loop diagrams. 

Because the effects of the running coupling gives only single
logarithmic terms it is enough to consider the remaining integrals 
to DL accuracy. 
Namely, we may trace  only the terms proportional to 
$\ln \frac{|p_1|^2}{s} \ln \frac{|p_2|^2}{s}$ from Fig.\ref{RGE}.
At the DL accuracy the spin structure of the amplitude is simple, so
that one  needs to consider the scalar integral only,  
\begin{eqnarray}
I=1+\frac{\alpha_s(\mu^2) C_F}{2\pi}(2p_1p_2)
\int \frac{d^4k}{(2\pi)^4} \frac{1}{(p_1+k)^2}
\frac{1}{(p_2-k)^2}\Big(1-\frac{\alpha_s\beta_0}{4\pi}
\ln (\frac{k^2}{\mu^2})\Big) \frac{i}{k^2}.
\end{eqnarray}
To evaluate this, we consider a slightly more general integral
\begin{eqnarray}
J=i\int \frac{d^4k}{(2\pi)^4} \frac1{(p_1+k)^2}
\frac{1}{(p_2-k)}\frac{\mu^{2\Delta}}{(k^2)^{1+\Delta}},
\end{eqnarray} 
which after expanding in $\Delta$ will give us the desired integral $I$. 
Using Feynman parameters, this  integral is reduced to 
\begin{eqnarray}
J=-\frac{1}{(4\pi)^2}\int\limits_{0}^{1}dy
\frac{1}{A\Delta}\mu^{2\Delta}\Big[ e^{-\Delta \ln (A+B)}- 
e^{-\Delta \ln (B)} \Big],
\end{eqnarray}
with
\begin{eqnarray}
A(y)=p_2^2y^2+2p_1p_2y+p_1^2,\\
B(y)=-2p_1p_2y-p_2^2y-p_1^2,\\
A(y)+B(y)=p_2^2y(-1+y).
\end{eqnarray}
The function $A(y)$ has two zeros, $y_\pm$: 
\begin{eqnarray}
A=p_2^2(y-y_+)(y-y_-),\quad
y_\pm=\frac{-2p_1p_2\pm\sqrt{(2p_1p_2)^2-4p_1^2p_2^2}
}{2p_2^2}.
\end{eqnarray}
For very small virtualities, $p_1^2, p_2^2 \to 0$, the roots are simplified to
$y_+=-\frac{p_1^2}{s}, y_-=-\frac{s}{p_2^2}$.
Expanding the integrand of $J$ in $\Delta$ up to second order we have 
\begin{eqnarray}
J&=&\frac{1}{(4\pi)^2}\int\limits_{0}^{1}dy\frac{1}{p_2^2(y-y_+)(y-y_-)}\mu^{2\Delta}
\Big( \ln\Big(\frac{p_2^2y(1-y)}{(2p_1p_2+p_1^2)y+p_1^2}\Big) \\
&+& \frac{\Delta}{2} \Big[ -\ln^2 (p_2^2y(1-y)) + \ln^2 ((2p_1p_2+p_2^2)y+p_1^2) \Big]
\Big).
\end{eqnarray}
The final integration over $y$ is simple, the result is
\begin{eqnarray}
J&=&\frac{1}{(4\pi)^2 2p_1p_2}
\Big( - \ln \frac{|p_1|^2}{s} \ln \frac{|p_2|^2}{s}+\frac{\Delta}{2} 
\ln \frac{|p_1|^2}{s} \ln \frac{|p_2|^2}{s} \ln
(\frac{|p_1^2||p_2^2|}{\mu^4})
\Big).
\end{eqnarray}
We see, that the first term in this equation reproduces the DL result 
from eq.(\ref{SudB}) and eq.(\ref{slresult}).  It can be checked,
that the second term suggests the  normalization of the
coupling constant to be, $\nu^2=\sqrt{|p_1^2||p_2^2|}$.  
Indeed, returning to the integral $I$, we find
\begin{eqnarray}
I=1-\frac{\alpha_s(\mu^2) C_F}{2\pi}
\ln \frac{|p_1|^2}{s} \ln \frac{|p_2|^2}{s}
\Big[ 1-\frac{\alpha_s(\mu^2)\beta_0}{4\pi}
\ln\Big(\frac{\sqrt{|p_1^2||p_2^2|}}{\mu^2}\Big) 
\Big].
\end{eqnarray}
It is clear that the last logarithm, containing the $\beta_0$ term,  
can be absorbed into the running coupling, giving $\alpha_s(\nu^2)$ with  
the normalization point $ \nu^2=\sqrt{|p_1^2||p_2^2|} $. 
 The exponentiation of the integral $I$ will give us the final off-shell
 Sudakov exponent, eq.(\ref{slresult}). 

%This result can be derived 
%in a variety of ways. One may perform a full two loop
%calculation, however, the essential result may be extracted using gauge 
%invariance and by considering the large $n_f$ limit. In this limit we evaluate 
%diagrams of the type shown in Fig.(3).   
%Essentially we need to consider the integral 
%We note that the same result is obtained if we 
%take the normalization of the coupling constant in the 
%un-integrated one loop amplitude to be $\mu^2=k^2_{\perp}$ \cite{Brodsky}.

In order to get single logarithms in eq.(\ref{slresult}) we have to include
the numerator and the spin structure. % more carefully. 
We do not present the 
details of these calculations here. Instead we note, that all logarithms 
we have accounted for are of infrared origin, $s \gg p_1^2,p_2^2\to 0$.  
We do not show the UV logarithms which come as a result of the 
renormalization of the quark mass.
Such terms can be omitted if the quark mass  
in the leading order result is normalized at a large scale $\mu^2=s$.   

The formula eq.(\ref{slresult}) reproduces the expression for the Sudakov 
form factor at non-equal virtualities at DL accuracy derived by  
Carrazone et. al.  in \cite{Carazzone:hj}, eq.(\ref{SudB}),
as well as at NLL with equal virtualities $p^2=p_1^2=p_2^2$ derived 
by Smilga in \cite{Smilga:uj}, eq.(\ref{Smilga}).

In addition, the normalization point 
$\nu^2=\sqrt{|p_1^2||p_2^2|}$ that we find 
reproduces that of the NLL results 
with equal virtualities $p^2=p_1^2=p_2^2$ derived by Smilga in 
\cite{Smilga:uj}, eq.(\ref{Smilga}).

This scale,  $\nu^2=\sqrt{|p_1^2||p_2^2|}$,  
has a very transparent origin. 
The vertex of the interaction of a soft gluon with an off-shell quark ($p_1^2$)
is described by the coupling $g(p_1^2)$. In the situation of  
gluon-exchange between two quarks with different virtualities, 
we have an effective  coupling  $g(p_1^2)g(p_2^2)$.  
 Using the running of the coupling $g^2(\mu^2)=4\pi\alpha_s(\mu^2)$, at one loop level, 
$\alpha(\mu^2)=\alpha_s(\nu^2)/(1+\frac{\alpha_s\beta_0}{4\pi}\ln(\frac{\mu^2}{\nu^2}))$,
we will find that the effective coupling   $g(p_1^2)g(p_2^2)$ is
reduced to $\alpha_s(\sqrt{|p_1^2 ||p_2^2| })$, which coincides with our previous results.

As a next step, we include this form factor inside the one loop  
triangle diagram 
and calculate the last loop integration with the  form
factor which now accounts for all large logarithms to NLL accuracy. 
The final result for the next-to-leading-logarithmic form 
factor reads
\begin{eqnarray}\label{rep1}
F^B=F^B_{DL}+F^B_{NLL},
\end{eqnarray}
with $F^B_{DL}$ from eq.(\ref{dlresult}) and 
\begin{eqnarray}\label{finalB}
F^B_{NLL}&=&\frac{1}{L}
F^{1-loop}_B\sum_{n=0}^{\infty} \frac{\Gamma (n+1)}{\Gamma (2n+2)}
(-\rho_B )^n\Big(3-\frac{\rho_B\beta_0}{C_F}\frac{n}{2n+2}\Big( 
\frac{n+1}{2(2n+3)}+\frac{\ln (s/\mu^2)}{L}\Big) \Big),   
\end{eqnarray} 
with $\beta_0=11-\frac{2n_f}3$, $n_f$ is the number of light flavors
\cite{Akhoury:2001mz}.

Topology C at NLL gives a slightly different result. 
In the previous section where we have already studied the DL result, we saw that the
``hard'' Sudakov form factor is developed as a result of the re-scattering of
a hard quark on the hard gluon.    
Let us start with the Sudakov form factor for the quark-gluon vertex 
with an off-shell outgoing quark of virtuality $p_1^2$ and a gluon of 
virtuality $p_2^2$. The result for the reduced form factor is 
\begin{eqnarray}
S_{NNL}&=&Exp \Big( 
-\frac{\alpha_s (\nu^2) C_A}{4\pi} \Big( 
\ln \frac{|p_1|^2}{s} \ln \frac{|p_2|^2}{s}
+\ln \frac{|p_1|^2}{s} 
+\frac{1}{2}\ln \frac{|p_2|^2}{s}   \Big) \\ \nonumber
&+& \frac{\alpha_s C_F}{4\pi}(\ln \frac{|p_1|^2}{s} -\frac{a_1}{4C_F}
\ln \frac{|p_2|^2}{s} \Big)\Big),
\end{eqnarray}
where the coefficient $a_1=\frac{10C_A}{3}-\frac{8T_Fn_F}{3}$ is related to the wave function
renormalization of the gluon.
Now we can obtain a resummed result by substituting this form factor in 
the known one loop integral of topology C. 
The DL result coincides with that of topology B (as we discussed 
previously, after accounting for the substitution $C_F \to C_A/2$).
The result for next-to-leading-logarithmic form factor reads 
\begin{eqnarray}\label{finalC}
F^C_{NLL}&=&
F^{1-loop}_C\sum_{n=0}^{\infty} \frac{\Gamma (n+1)}{\Gamma (2n+2)}
(-\rho_C)^n\Big( 3-\frac{2C_F}{C_A}+\frac{a_1}{2C_A} \\ \nonumber 
&-& \frac{2\rho_C\beta_0}{C_A}\frac{n}{2n+2}\Big( 
\frac{n+1}{2(2n+3)}+\frac{\ln (s/\mu^2)}{L}\Big) \Big)   
\end{eqnarray} 
The final result for the amplitude is eq.(\ref{ampNLL}) together with 
eq.(\ref{rep1},\ref{finalB},\ref{finalC}).
That concludes our derivation of the NLL form factors.

We emphasize that throughout this subsection we have taken into account all logarithms of the
Sudakov type only. In section 4 we argue that to the NLL level this is justified because of
cancellations between diagrams that is similar to the DL case. In other words, logarithms of
the NLL type coming from other sources (than the Sudakov form factors) cancel between 
contributions of subsets of diagrams.

\section{Cancellation mechanisms at the leading logarithmic level}

The simple expressions for the resummed results given in the previous
section are a consequence of the cancellation of double logarithmic terms 
which are not of the Sudakov type. In fact for the exponentiation to work there
has to be a large number of cancellations between subgroups of diagrams.
Such cancellations have been discussed in the literature explicitly at the 
two loop \cite{Fadin:1997sn} and the three loop levels
\cite{Melles:1998gu}. We agree
with the results of these papers however the new element we would like to
add here is to point out the general mechanism which is responsible for 
these cancellations. We discuss this mechanism on various examples and 
write down identities which guarantee the cancellations to higher loop
orders.

\subsection{The dipole mechanism for cancellation}

\begin{figure}
\centerline{\epsfig{file=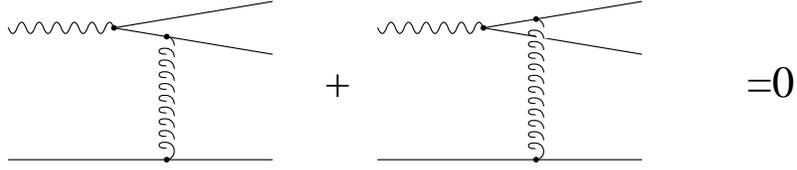,scale=0.85}}
\caption{\label{generic}
The general mechanism of the DL cancellation in the groups 1 and 2 of section 3.2
is the  dipole interaction of a collinear pair of quarks with one soft gluon.}
\end{figure}

The relevant cancellations at the double logarithmic level are due to 
a single mechanism - the dipole mechanism. Two fermions with opposite
charges form a dipole when separated by a short distance.
When observing from far away, or probing by means of a gluon with long
wavelength, one finds the dipole system to be neutral to leading
order.
To be more specific, the total amplitude for exchanging a soft gluon with
two such oppositely charged fermions, as shown by the sum of the two generic
Feynman diagrams in Fig. (\ref{generic}), vanishes in the infrared
sensitive region (double and single logarithms). This infrared sensitive
region is that of a soft gluon, i.e., a gluon with all components of the
4-momentum small,
$k^{\mu}=(k^+,k^-,k_{\bot})$,
\begin{displaymath}
k^+,k^-,k_{\bot} \sim \lambda \rightarrow 0,
\end{displaymath}
This soft region is responsible for the double logarithms.

\begin{figure}
\centerline{\epsfig{file=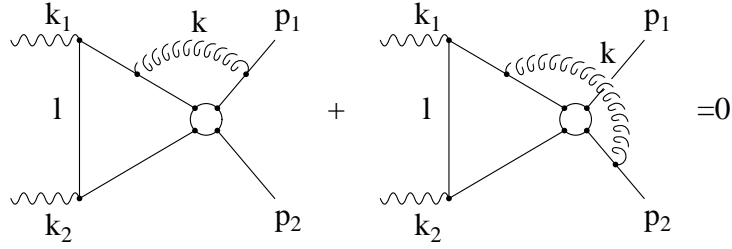,scale=0.85}}
\caption{\label{mixed}
Example of a cancellation dictated by the dipole mechanism.}
\end{figure}

The cancellation can be shown explicitly by employing the Grammar-Yennie
\cite{Sterman:Book} decomposition and the Ward identity.To illustrate the
mechanism let us consider a special case shown in, Fig. (\ref{mixed}).
The amplitude for the first diagram is
\begin{equation}
M_{1}=\Gamma_{1}\frac{1}{(\slk_{1}-\sll)-m}\gamma^{\beta}
	\frac{1}{(\slk_{1}-\sll-\slk)-m}\Gamma_{2}
	\frac{1}{(\slp_{1}+\slk)-m}\gamma^{\alpha}
	u(p_{1})D_{\alpha\beta}(k),
\end{equation}
where we have explicitly written down only the factors relevant for the
cancellation and denoted the rest of the corresponding expressions for
the diagram by $\Gamma_1$ and
$\Gamma_2$. Their explicit form are
of no interest as long as they are common to both the diagrams, $M_1$ and $M_2$.
Employing the Grammar-Yennie decomposition we write for the gauge boson
propagator,
\begin{eqnarray}
D_{\alpha \beta}(k)=G_{\alpha \beta}+K_{\alpha \beta}(k), \\
G_{\alpha \beta}(k)=D_{\alpha \beta}-
	\frac{k_{\alpha}D_{\beta \tau}p_1^{\tau}}{k \cdot p_1}, \\
K_{\alpha \beta}=\frac{k_{\alpha}D_{\beta \tau}p_1^{\tau}}{k \cdot p_1}, \\
\end{eqnarray}
and keep only the term $K_{\alpha \beta}(k)$, i.e., we make the replacement
\begin{equation}
D_{\alpha\beta}(k)=\frac{g_{\alpha\beta}}{k^2} \rightarrow
	\frac{k_{\alpha}g_{\beta\tau}p_{1}^{\tau}}{k^2(p_1 \cdot k)},
\end{equation}
which leaves the infrared behavior unchanged \cite{Sterman:Book}, we obtain
\begin{eqnarray}
M_1 &\rightarrow& \Gamma_{1} \frac{\slk_1}{(k_1-l)^2-m^2} \gamma^{\beta}
	\frac{\slk_1}{(k_1-l-k)^2-m^2} \Gamma_2
        \frac{1}{(\slp_{1}+\slk)-m}\gamma^{\alpha}u(p_1)
	\frac{k_{\alpha}g_{\beta\tau}p_{1}^{\tau}}{k^2(p_1 \cdot k)}
\\
&=& -\Gamma_{1} \frac{\slk_1}{(k_1-l)^2-m^2} \slp_1
        \frac{\slk_1}{(k_1-l-k)^2-m^2} \Gamma_2 u(p_1)
	\frac{1}{k^2(p_1 \cdot k)}
\\
&=& - \Gamma_1
	\frac{2 \slk_1(p_1 \cdot k_1)}{[(k_1-l)^2-m^2][(k_1-l-k)^2-m^2]}
	\Gamma_2 u(p_1) \frac{1}{k^2(p_1 \cdot k)},
\end{eqnarray}
where we have applied the Ward identity on the outgoing fermion line.
Summing the two diagrams together, we have
\begin{equation}
M_1+M_2 = M \left[ \frac{p_1 \cdot k_1}{p_1 \cdot k}-
	\frac{p_2 \cdot k_2}{p_2 \cdot k}\right],
\end{equation}
where the relative minus sign arises from the fact that the soft gluon
interacts with two oppositely charged fermions and
\begin{displaymath}
M = - \Gamma_1  
        \frac{\slk_1}{[(k_1-l)^2-m^2][(k_1-l-k)^2-m^2]}
        \Gamma_2 u(p_1) {1 \over k^2}.
\end{displaymath}

In order to prove the sum, $M_1+M_2$, vanishes, we use Sudakov
parameterization,
\begin{equation}
k = \beta k_1+\alpha k_2 + k_{\bot},
\end{equation}
and proceed with the soft approximation,
\begin{equation}
s \alpha, s \beta \gg k_{\bot}^2.
\end{equation}
We arrive at
\begin{equation}
\left[ \frac{p_1 \cdot k_1}{p_1 \cdot k} -
\frac{p_2 \cdot k_2}{p_2 \cdot k}\right]
= \frac{s}{2}
\left[\frac{1}{(\beta+\alpha)\frac{s}{2}-p_{1\bot} \cdot k_{\bot}} -
	\frac{1}{(\beta+\alpha)\frac{s}{2}-p_{2\bot} \cdot k_{\bot}}
\right].
\end{equation}
Again, in the soft approximation, $p_{1\bot} \cdot k_{\bot}$
and $p_{2\bot} \cdot k_{\bot}$ are negligible compared to
$(\beta + \alpha)\frac{s}{2}$ due to the $\delta$-function
\begin{equation}
\delta(s \alpha \beta -k_{\bot}^2 - \lambda^2)
\end{equation}
arising from the pole of the gauge boson propagator. This can be readily
seen in the expansion
\begin{equation}
\frac{1}{(\beta + \alpha) \frac{s}{2}-p_{1\bot} \cdot k_{\bot}} =
	\frac{1}{(\beta + \alpha)\frac{s}{2}}
	\left\{ 1+ \left[ \frac{p_{1\bot} \cdot k_{\bot}}{(\beta +\alpha)
	\frac{s}{2}} \right]^2
	+\left[\frac{p_{1\bot} \cdot k_{\bot}}{(\beta +
\alpha)\frac{s}{2}}\right]^4+\cdots\right\},
\end{equation}
since the odd terms in $p_{1\bot} \cdot k_{\bot}$ vanish when integrating
over the angle between $p_{1\bot}$ and $k_{\bot}$.
Further,
\begin{equation}
\left[ \frac{p_{1\bot} \cdot k_{\bot}}{(\beta + \alpha)\frac{s}{2}}
\right]^{2n} =
\left\{
\begin{array}{ll}
2^n \left( \frac{\beta}{\alpha} \right)^n \frac{p_{1\bot}^{2n}}{s^n}
\cos^{2n}\theta & \alpha \gg \beta \\ \\
2^n \left[ \frac{\alpha \beta}{(\alpha+\beta)^2} \right]^n 
\frac{p_{1\bot}^{2n}}{s^n} \cos^{2n}\theta & \alpha \sim \beta \\ \\
2^n \left( \frac{\alpha}{\beta} \right)^n \frac{p_{1\bot}^{2n}}{s^n}
\cos^{2n}\theta & \beta \gg \alpha
\end{array}
\right.
\end{equation}
and power suppression arises from the term
$\left( \frac{\beta}{\alpha} \right)^n$ with $\alpha \gg \beta$,
$\left( \frac{\alpha}{\beta} \right)^n$ with $\beta \gg \alpha$ and
$\left( \frac{p_{1\bot}^2}{s} \right)^n$ for all the three cases.

Thus we have seen that the two diagrams indeed cancel each other in the double
logarithmic approximation. The explicit calculation is applicable to the
generic diagrams in Fig. (\ref{mixed}) if we recall that what really matters
here is nothing but the exchange of the soft gluon with the
fermion-antifermion pair.
The same conclusion holds for a exchanging collinear gluon in the single
logarithm approximation as we will discuss later.

\subsection{Three loop examples}

\begin{figure}
\centerline{\epsfig{file=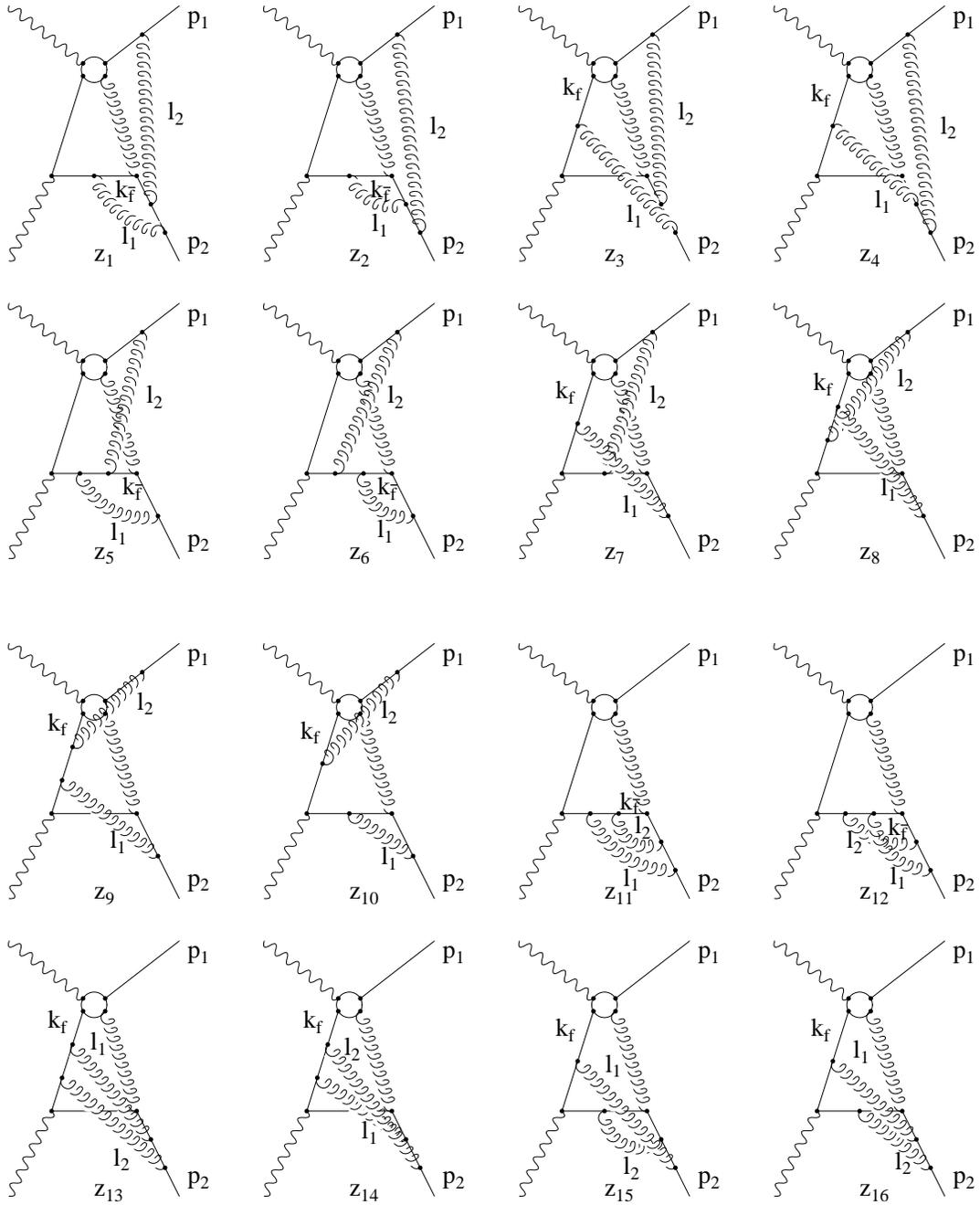,scale=0.75}}
\caption{\label{TopC}
The Abelian diagrams of the topology C.}
\end{figure}

We will next discuss how the dipole mechanism for cancellation works in 
certain 3 loop
examples. More specifically, we will show the cancellation of double 
logarithms in the diagrams $z_{1}-z_{16}$ of topology C. For simplicity 
of presentation we first discuss only the Abelian case omitting all
group theory factors
and in the next section the extension to the non-Abelian case will be 
presented. These diagrams can be grouped in twos or threes
according to the cancellation as seen below.
\begin{enumerate}
\item $z_1+z_2 = 0$
\item $z_2+z_4 = 0$
\item $z_5+z_6+z_7 = 0$
\item $z_8+z_9+z_{10} = 0$
\item $z_{11}+z_{12}+z_{15} = 0$
\item  $z_{13}+z_{14}+z_{16} = 0$
\end{enumerate}
These six groups cover all the Abelian-like diagrams of Fig. \ref{TopC}. 

\par
{\bf Group 1.}\\

We start with group 1.
The amplitudes $z_1$ and $z_3$ may be written as 
\begin{eqnarray}
z_1=\int \frac{d k_f}{(2\pi)^4}\int\frac{dl_1}{(2\pi)^4}\Big[\sle 
\frac{i}{\slk_{\bar f} +\sll_1 -m_q+i0}i\gamma^{\mu}
\Big] M^{(1)}_{\mu}
\end{eqnarray}
\begin{eqnarray}
z_3=\int \frac{d
k_f}{(2\pi)^4}\int\frac{dl_1}{(2\pi)^4}\Big[i\gamma^{\mu} 
\frac{i}{\slk_{f} - \sll_1 - m_q+i0}\sle
\Big] M^{(1)}_{\mu}
\end{eqnarray}
where $M^{(1)}_{\mu}$ is some one loop sub-diagram, 
which is identical in both $z_1$ and $z_3$. 
Such a representation of the diagrams is very convenient because 
it turns out that the cancellation of double logarithmic contributions 
could be traced without explicit computations of the separate diagrams. 
Instead we observe the cancellations by comparing diagrams in some 
subgroups and using kinematic simplifications specific to the double 
logarithmic approximation.
 
Because the final quark line is hard
we might write $M^{(1)}_{\mu}= p_{2\mu} M^1$. 
The momenta $l_1$ must be softer than $k_{\bar f}$, otherwise no double
logarithms can appear. Then we find that integrand of  
the diagram $z_1$ contains $(p_2 \cdot k_{\bar f})$. Therefore 
decomposing the momenta  $(k_{\bar f})$ into $k_f$ and $p_2$ we find that 
only the component parallel to $k_f$ contributes. We remind the reader that 
$p_2^2=m^2_q \to 0$. Introducing $n_\mu$, the unit vector parallel to $k_f$, 
\begin{equation}
n^\mu=\frac{k_f^\mu}{|k_f|},
\end{equation}
we obtain for the sum of the diagrams
\begin{eqnarray}
z_1+z_3=-\int \frac{d k_f}{(2\pi)^4}\int\frac{dl_1}{(2\pi)^4}
\sle M^1 (n \cdot p_2) 
\Big[ \frac{1}{nl_1+i0}+\frac{1}{-nl_1+i0}\Big] M^1=0
\end{eqnarray}
Thus the leading contribution in the integrand, responsible for the
double logarithmic asymptotic, cancels. This is shown in Fig. \ref{generic}.

As we will see below the mechanism of the cancellation is very general
and is related to the dipole mechanism. 
Indeed we may define an impact factor as a sub-diagram, where a
real photon produces a nearly on-shell quark and antiquark, which interacts 
with the soft gluons. The remarkable fact is that in the configuration responsible for double logarithms
both particles move in the same direction. Because the quark momenta is harder than the gluon momenta, 
only the direction of its movement plays a role. 
 $$
\frac{k_{\bar f} p}{k_{\bar f} l}=\frac{n p}{n l}.  
$$
Further, because the $b\bar b$ system  does not have a net color charge, being produced by
a photon, the first term in the multi-pole expansion of interaction of this
system with any number of soft gluons is the color-dipole moment. 
This leads to the power suppression of the sub-diagram and the loss of the double logarithms.

 {\bf Group 2.}\\
We continue with group 2 which constitutes diagrams $z_2$ and $z_4$.
The corresponding amplitudes could be written as 
\begin{eqnarray}
z_2=\int \frac{d k_f}{(2\pi)^4}\int\frac{dl_1}{(2\pi)^4}\Big[\sle 
\frac{i}{\slk_{\bar f} +\sll_1 -m_q+i0}i\gamma^{\mu}
\Big] M^{(2)}_{\mu}
\end{eqnarray}
\begin{eqnarray}
z_4=\int \frac{d
k_f}{(2\pi)^4}\int\frac{dl_1}{(2\pi)^4}\Big[i\gamma^{\mu} 
\frac{i}{\slk_{f} - \sll_1 - m_q+i0}\sle
\Big] M^{(2)}_{\mu}
\end{eqnarray}
where as before, $M^{(2)}_{\mu}$ is some  sub-diagram,and we may write 
$M^{(2)}_{\mu} = p_{2\mu} M^2$. 
Again, the momenta $l_1$ is softer than $k_{\bar f}$ and 
only the component of $k_{\bar f}$  parallel to $k_f$ contributes. 
We have
\begin{eqnarray}
z_2+z_4=-\int \frac{d k_f}{(2\pi)^4}\int\frac{dl_1}{(2\pi)^4}
\sle M^{(2)} (n \cdot p_2) 
\Big[ \frac{1}{nl_1+i0}+\frac{1}{-nl_1+i0}\Big] M^2=0
\end{eqnarray}
We thus see that the leading contribution to the integrand, which gives rise to the
double logarithmic asymptotic, cancels.

 {\bf Group 3.}\\
We turn now to the amplitudes for the diagrams $z_5, z_6, z_7$ which may be written as 
\begin{eqnarray}
z_5=\int \frac{d k_f}{(2\pi)^4}\int\frac{dl_1}{(2\pi)^4}
\int\frac{dl_2}{(2\pi)^4}
\Big[
\sle \frac{i}{\slk_{\bar f} +\sll_1 +\sll_2 -m_q+i0}i\gamma^{\mu}
\frac{i}{\slk_{\bar f} +\sll_2 -m_q+i0}i\gamma^{\nu}
\Big] M^{(5)}_{\mu,\nu}
\end{eqnarray}
\begin{eqnarray}
z_6=\int\frac{dk_f}{(2\pi)^4}\int\frac{dl_1}{(2\pi)^4}
\int\frac{dl_2}{(2\pi)^4}\Big[
\sle\frac{i}{\slk_{\bar f} +\sll_1 +\sll_2 -m_q+i0}i\gamma^{\nu}
\frac{i}{\slk_{\bar f} +\sll_1 -m_q+i0}i\gamma^{\mu}
\Big] M^{(6)}_{\mu,\nu}
\end{eqnarray}
\begin{eqnarray}
z_7=\int\frac{d k_f}{(2\pi)^4}\int\frac{dl_1}{(2\pi)^4}\int\frac{dl_2}{(2\pi)^4}\Big[
i\gamma^{\mu}\frac{i}{\slk_{f} -\sll_1 -m_q+i0}\sle 
\frac{i}{\slk_{\bar f} +\sll_2 -m_q+i0}i\gamma^{\nu}
\Big] M^{(7)}_{\mu,\nu}
\end{eqnarray}
it is clear from the diagrams that we can write to the accuracy needed
for the relevant subgraphs,
$M^{(5)}_{\mu,\nu}= M^{(6)}_{\mu,\nu}=M^{(7)}_{\mu,\nu}=p_{2\mu} 
p_{1\nu}M^{3}$.
The momenta $l_1,l_2$ are softer than $k_{\bar f}$ and 
only the component of $k_{\bar f}$  parallel to $k_f$ contributes. 
After simplifications, we obtain
\begin{eqnarray}
z_5+z_6+z_7&=&\int \frac{d k_f}{(2\pi)^4}\int\frac{dl_1}{(2\pi)^4}\int\frac{dl_2}{(2\pi)^4}
\sle M^{3} (n \cdot p_2)(n \cdot p_1) \\ \nonumber
&\Big[& 
\frac{1}{n(l_1+l_2)}\frac{1}{(nl_2)}
+\frac{1}{n(l_1+l_2)}\frac{1}{(nl_1)}
- \frac{1}{(nl_1)}\frac{1}{(nl_2)} 
\Big] =0
\end{eqnarray}
Again, we see that the leading contribution to the integrand, 
which could give rise to the double logarithmic asymptotic, cancels.

\begin{figure}
\centerline{\epsfig{file=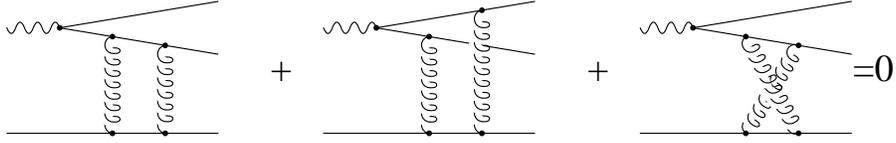,scale=0.85}}
\caption{\label{fig2}
The general mechanism of the DL cancellation in the groups 3-6 is the 
dipole interaction of a collinear pair of quarks with two soft gluons.}
\end{figure}

  {\bf Group 4.}\\
The  amplitudes for $z_8, z_9, z_{10}$ could be written as 
\begin{eqnarray}
z_8=\int \frac{d k_f}{(2\pi)^4}\int\frac{dl_1}{(2\pi)^4}\int\frac{dl_2}{(2\pi)^4}\Big[
i\gamma^{\mu}\frac{i}{\slk_{f} - \sll_1 -m_q+i0}i\gamma^{\nu}
\frac{i}{\slk_{f} -\sll_1-\sll_2 -m_q+i0}\sle
\Big] M^{(8)}_{\mu,\nu}
\end{eqnarray}
\begin{eqnarray}
z_9=\int \frac{d k_f}{(2\pi)^4}\int\frac{dl_1}{(2\pi)^4}\int\frac{dl_2}{(2\pi)^4}\Big[
i\gamma^{\nu}\frac{i}{\slk_{f} - \sll_2 -m_q+i0}i\gamma^{\mu}
\frac{i}{\slk_{f} -\sll_1-\sll_2 -m_q+i0} \sle
\Big] M^{(9)}_{\mu,\nu}
\end{eqnarray}
\begin{eqnarray}
z_{10}=\int \frac{d k_f}{(2\pi)^4}\int\frac{dl_1}{(2\pi)^4}\int\frac{dl_2}{(2\pi)^4}\Big[
i\gamma^{\nu}\frac{i}{\slk_{f} - \sll_2 -m_q+i0}\sle
\frac{i}{\slk_{\bar f} + \sll_1-m_q + i0} i\gamma_{\mu}
\Big] M^{(10)}_{\mu,\nu}
\end{eqnarray}
with  $M^{(8)}_{\mu,\nu}=
M^{(9)}_{\mu,\nu}=M^{(10)}_{\mu,\nu}=p_{2\mu} p_{1\nu} 
M^{4}$ to the desired accuracy for the sub-diagrams. 
The momenta $l_1,l_2$ are softer than $k_{\bar f}$ and 
only the component of $k_{\bar f}$  parallel to $k_f$ contributes. 
After simplifications, we obtain
\begin{eqnarray}
z_8+z_9+z_{10}&=&
\int \frac{d k_f}{(2\pi)^4}\int\frac{dl_1}{(2\pi)^4}\int\frac{dl_2}{(2\pi)^4}
\sle M^{4} (n \cdot p_2)(n \cdot p_1) \\ \nonumber
&\Big[& - \frac{1}{(nl_1)}\frac{1}{(nl_2)}
+\frac{1}{n(l_1+l_2)}\frac{1}{(nl_2)}
+\frac{1}{n(l_1+l_2)}\frac{1}{(nl_1)}
   \Big] =0
\end{eqnarray}
We see that the leading contribution to the integrand cancels.
The mechanism of the cancellation is identical to the one of group 3.
It is important to note specially for purposes of the next section that
$M^{4}=M^{3}$. Hence one might find
another 2 sets of 3 diagrams in groups 3 and 4, 
which cancel in groups of three.

{\bf Group 5.}\\
The  amplitudes for $z_{11}, z_{12}, z_{15}$ could be written as 
\begin{eqnarray}
z_{15}=\int \frac{d k_f}{(2\pi)^4}\int\frac{dl_1}{(2\pi)^4}
\int\frac{dl_2}{(2\pi)^4}
\Big[
i\gamma^\mu \frac{i}{\slk_{f} - \sll_1 -m_q+i0}\sle
\frac{i}{\slk_{\bar f} +\sll_2 -m_q+i0} i\gamma^{\nu}
\Big] M^{(15)}_{\mu,\nu}
\end{eqnarray}
\begin{eqnarray}
z_{12}=\int \frac{d k_f}{(2\pi)^4}\int\frac{dl_1}{(2\pi)^4}
\int\frac{dl_2}{(2\pi)^4}
\Big[
\sle \frac{i}{\slk_{\bar f} + \sll_1+\sll_2 -m_q+i0}i\gamma^{\mu}
\frac{i}{\slk_{\bar f} +\sll_2 -m_q+i0}i\gamma^{\nu} 
\Big] M^{(12)}_{\mu,\nu}
\end{eqnarray}
\begin{eqnarray}
z_{11}=\int \frac{d k_f}{(2\pi)^4}\int\frac{dl_1}{(2\pi)^4}
\int\frac{dl_2}{(2\pi)^4}
\Big[
\sle \frac{i}{\slk_{\bar f} + \sll_1+\sll_2 -m_q+i0}i\gamma^{\nu}
\frac{i}{\slk_{\bar f} +\sll_1 -m_q+i0}i\gamma^{\nu} 
\Big] M^{(11)}_{\mu,\nu}
\end{eqnarray}
with  $M^{(11)}_{\mu,\nu}=
M^{(12)}_{\mu,\nu}=M^{(15)}_{\mu,\nu}= p_{1\mu} p_{1\nu} 
M^{5}$ the relevant sub-diagrams.
The momenta $l_1,l_2$ have to be softer than $k_{\bar f}$ and 
only  component of $k_{\bar f}$  parallel to $k_f$ contributes. 
After simplifications, we obtain
\begin{eqnarray}
z_{11}+z_{12}+z_{15}&=&
\int \frac{d k_f}{(2\pi)^4}\int\frac{dl_1}{(2\pi)^4}\int\frac{dl_2}{(2\pi)^4}
\sle M^{5} (n \cdot p_1)(n \cdot p_1) \\ \nonumber
&\Big[& - \frac{1}{(nl_1)}\frac{1}{(nl_2)}
+\frac{1}{n(l_1+l_2)}\frac{1}{(nl_2)}
+\frac{1}{n(l_1+l_2)}\frac{1}{(nl_1)}
   \Big] =0
\end{eqnarray}
We again observe the cancellation of the leading contribution to the integrand.
The mechanism of cancellation is identical to the one of group 3 and 
is the just the dipole mechanism.

{\bf Group 6.}\\
Finally the  amplitudes for $z_{13}, z_{14}, z_{16}$ 
could be written as 
\begin{eqnarray}
z_{16}=\int \frac{d k_f}{(2\pi)^4}\int\frac{dl_1}{(2\pi)^4}
\int\frac{dl_2}{(2\pi)^4}
\Big[
i\gamma^\mu \frac{i}{\slk_{f} - \sll_1 -m_q+i0} \sle 
\frac{i}{\slk_{\bar f} +\sll_2 -m_q+i0} i\gamma^{\nu}
\Big] M^{(16)}_{\mu,\nu}
\end{eqnarray}
\begin{eqnarray}
z_{14}=\int \frac{d k_f}{(2\pi)^4}\int\frac{dl_1}{(2\pi)^4}
\int\frac{dl_2}{(2\pi)^4}
\Big[
i\gamma^{\mu}\frac{i}{\slk_{f} -\sll_2 -m_q+i0}i\gamma^{\nu}
\frac{i}{\slk_{ f} -\sll_1- \sll_2 -m_q+i0}\sle  
\Big] M^{(14)}_{\mu,\nu}
\end{eqnarray}
\begin{eqnarray}
z_{13}=\int \frac{d k_f}{(2\pi)^4}\int\frac{dl_1}{(2\pi)^4}
\int\frac{dl_2}{(2\pi)^4}
\Big[
i\gamma^{\nu}\frac{i}{\slk_{f} -\sll_1 -m_q+i0}i\gamma^{\mu}
\frac{i}{\slk_{ f} -\sll_1- \sll_2 -m_q+i0} \sle  
\Big] M^{(13)}_{\mu,\nu}
\end{eqnarray}

where as before we write to this accuracy, $M^{(13)}_{\mu,\nu}=
M^{(14)}_{\mu,\nu}=M^{(16)}_{\mu,\nu}= p_{1\mu} p_{1\nu} 
M^{6}$.
The momenta $l_1,l_2$ have to be softer than $k_{\bar f}$ and 
only the  component of $k_{\bar f}$  parallel to $k_f$ contributes. 
After simplifications, we obtain
\begin{eqnarray}
z_{13}+z_{14}+z_{16}&=&
\int \frac{d k_f}{(2\pi)^4}\int\frac{dl_1}{(2\pi)^4} \int\frac{dl_2}{(2\pi)^4}
\sle M^{6} (n \cdot p_1)(n \cdot p_1) \\ \nonumber
&\Big[& - \frac{1}{(nl_1)}\frac{1}{(nl_2)}
+\frac{1}{n(l_1+l_2)}\frac{1}{(nl_2)}
+\frac{1}{n(l_1+l_2)}\frac{1}{(nl_1)}
   \Big] =0
\end{eqnarray}
which explicitly shows the cancellation of the leading terms.
The mechanism of the cancellation is identical to the one for the
earlier groups and 
is the dipole mechanism.
Note that for these two groups as well, 
$M^{5}=M^{6}$ after interchanging $l_1$ and
$l_2$. 
Hence one might expect to find another two sets of three diagrams in
groups 5 and 6, 
which cancel in groups of three.

In conclusion, we have shown that 
all diagrams $z_1-z_{16}$ cancel in group of two or three due to the 
dipole mechanism.

\subsection{Inclusion of non-abelian diagrams}

The inclusion of the non-abelian contributions do not pose any new 
difficulties. In fact,
there are many simplifications because the final state quark and
antiquark are produced by 
photons and hence carry no net color charge. When the color factors 
are included the cancellations
take place between diagrams with the same such factors. For example, 
consider the cancellation
between the diagrams of group 1. Here $z_1$ and $z_3$ have the same
group  factor since the only 
difference between them is an abelian vertex. The same is true for group
2. Thus the cancellation between
the diagrams of groups 1 and 2 proceed as in the previous section even
in the full non-abelian theory. 

Consider next, group 3 of the previous subsection. 
For this case we will discuss the cancellation in two different ways. In
the first method we note
that only the group theory factor for the diagram $z_6$ is different
from that of $z_5$ and $z_7$. 
Explicitly, for $z_6$ the factor is $T_a T_c T_a T_b T_c T_b$ while for the 
other two it is
$T_aT_cT_bT_aT_cT_b$. All other factors are the same as in the previous 
section with the rule that
the group factors are overall multiplicative. In particular, the
sub-diagrams $M^{(5)}_{\mu,\nu}=
M^{(6)}_{\mu,\nu}=M^{(7)}_{\mu,\nu}=p_{2\mu} p_{1\nu}M^{3}$ contain no 
color matrices. Now we can write
the color factor of $z_6$ as the one for $z_5$ and $z_7$ plus a left
over term which is
$if_{abc}T_aT_cT_b(C_F-C_A/2)$. Consider now the diagrams in group 4 of
the previous section. Here
the group factor associated with $z_8$ is $T_aT_cT_bT_aT_cT_b$ which is 
different from $z_9$ and
$z_{10}$ which is $T_aT_cT_aT_bT_cT_b$. The difference now is
$-if_{abc}T_aT_cT_b(C_F-C_A/2)$. Using the fact that $M^3=M^4$ we see
that the left over pieces 
from groups 3 and 4 cancel each other out. It is easy to check that the
same applies to the combination of groups 5 and 6.

It is clear from the above discussion, that it is much more
straightforward in the non abelian case
to change the assignment of the diagrams into different groups. It is
easy to see that apart from
the group theory factors, and in the soft approximation, the diagrammatic
expressions for $z_6$ and $z_8$
of the previous subsection (from groups 3 and 4) are the same (using 
of course $M^3=M^4$). From groups 5
and 6 of the previous subsection the same applies to the expressions 
for $z_{12}$ and $z_{14}$ (in this
case using $M^5=M^6$). Thus the following assignment of the diagrams in 
the non-abelian case into
different groups will make sure that all diagrams in a group have the 
same group theoretical factors:
\begin{enumerate}
\item $z_1+z_3 = 0$
\item $z_2+z_4 = 0$
\item $z_5+z_8+z_7 = 0$
\item $z_6+z_9+z_{10} = 0$
\item $z_{11}+z_{14}+z_{15} = 0$
\item  $z_{13}+z_{12}+z_{16} = 0$
\end{enumerate}
The new assignment does not change the results for the abelian case and
now the cancellation in the
non-abelian case proceeds within each group. The only non-trivial result one
must use is the equalities
$M^3=M^4$ and $M^5=M^6$. These are always seen to be true in the soft 
approximation. In the present
grouping, the diagrams in each group are seen to be related to each
other by a cyclic permutation
of the gluon lines. Such a cyclic permutation leaves the color factor 
unchanged.

\begin{figure}
\centerline{\epsfig{file=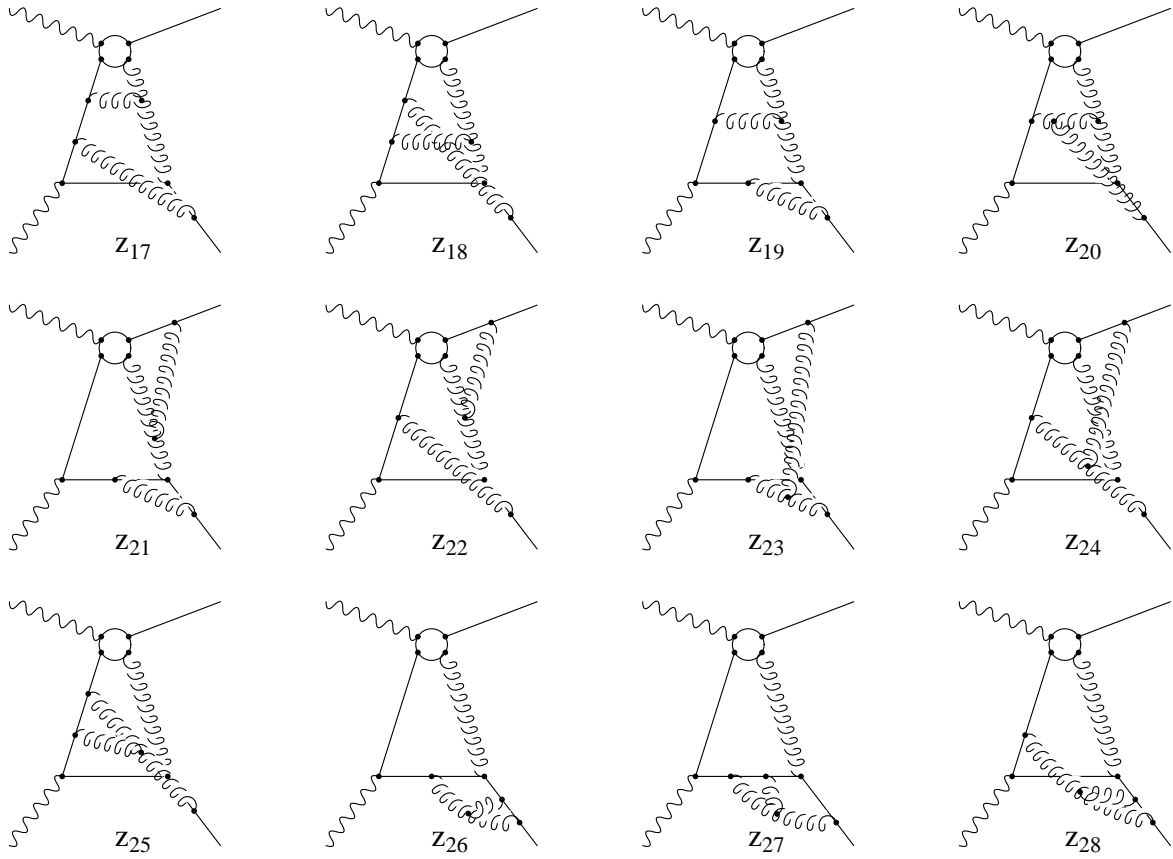, scale=0.85}
}
\caption{\label{nonab}
Non-Abelian diagrams.
}
\end{figure}

The cancellation between the other non Abelian diagrams shown in 
Fig.\ref{nonab} also proceeds similarly:
For example, $z_{21}$ cancels $z_{22}$; $z_{23}$ cancels $z_{24}$ and so on. 
 
This discussion now has been set up for generalization to all orders.

\subsection{Generalization to higher orders}

We see that the cancellation at the 3 loop level discussed in the previous subsection relies
on the following:
(1) the soft approximation,(2) algebraic identities like
\begin{eqnarray} \label{id_DL}
{1 \over -p \cdot k_1} + {1 \over p \cdot k_1}=0 \\
{1 \over -p \cdot k_1}{1 \over -p \cdot (k_1+k_2)}+ 
{1 \over p \cdot k_1}{1 \over -p \cdot k_2}+ 
{1 \over p \cdot (k_1+k_2)}{1 \over p \cdot k_2} = 0, \nonumber
\end{eqnarray}
and that diagrams in the same group are related by a cyclic permutation
of the gluon lines. In the
above, $p$ is a generic hard momentum and $k_i$ are the soft ones.
The soft approximation essentially tells us that a soft gluon does not
see spin and more explicitly
if $p$ generically denotes a hard fermion momentum then we can
consistently use,${i \over 2p \cdot k}$
for the hard propagator and $2p^{\mu}$ for the vertex factors. This will
ensure the equalities like
$M^3=M^4$ needed for the cancellations to occur for our process. As far
as the algebraic identities are
concerned, they are in fact special cases of the Sterman-Libby 
identities \cite{Libby:nr}. These
identities obtained by considering diagrams related by cyclic
permutations of the gluon lines read in
general:
\begin{equation}
\sum_{m=0}^{n}\prod_{i=0}^{m-1}{1 \over p \cdot q_m -p \cdot q_i}
\prod_{j=m+1}^n{1 \over p \cdot q_m -p \cdot q_j}=0
\end{equation}
In the above, $q_i \equiv k_1+k_2+.....+k_i$ and $q_{i=0} \equiv 0$. These 
identities are easily 
seen to reduce to Eqs.(\ref{id_DL}) for the cases $n=1,2$.
Thus we see each of the technical ingredients have a generalization to
higher loop orders. 
The physics of the dipole mechanism and the color singlet nature of the 
final (and initial)
states combined with the above guarantee the cancellations needed to all
orders.
\begin{figure}
\centerline{\epsfig{file=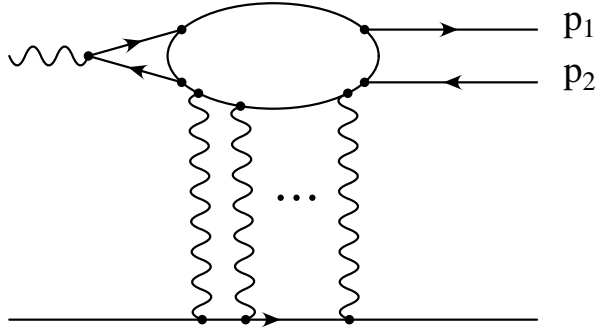,scale=0.85}}
\caption{\label{fig4}
The general mechanism of the DL cancellation is the 
dipole interaction of a collinear pair of quarks with many soft gluons.}
\end{figure}

\section{Contributing diagrams at the next-to-leading logarithmic order}

The goal of this section is to identify those diagrams that either
vanish or cancel some other diagrams at next-to-leading logarithmic
order. The diagrams left over are then just those needed in the
discussion in Section 2.3. Our explicit analysis is only at the two loop
level and at the end we argue that the results hold to NLL accuracy.
In the first subsection, we discuss the regime contributing to the
next-to-leading logarithms. We then introduce a power counting
technique that enables us to discover the non-vanishing diagrams and
make appropriate approximations.
Armed with these two techniques, we are able to exclude a set of diagrams
without actually carrying out the loop integrations.
We remind the reader that we work throughout in
the Feynman gauge.
In this gauge we will see that for the process under consideration, the
additional logarithms at the NLL level must be of collinear origin.

\subsection{Sources of single logarithms}

In a two-particle scattering process, all momenta lie in the same plane.
We can take two of
the independent momenta, $k_1$ and $k_2$, as '+' and '-' direction, and
denote the third momentum as $p$ ($p$ could be either $p_1$ or $p_2$).
The loop momentum $l$ can be expressed in terms of light cone variables
\begin{equation}
l = \alpha k_1 + \beta k_2 + l_{\perp}.
\end{equation}

We note that the following integral in the soft regime
\begin{equation}
\int d\alpha d\beta \frac{1}{\alpha^i} \frac{1}{\beta^j}
	\Theta(s \alpha \beta - m^2), i, j = ..., -2, -1, 0, 1, 2, ...,
\end{equation}
gives terms like $\log^2 \frac{s}{m^2}$, $(\frac{m^2}{s})^n \log \frac{s}{m^2}$
or $(\frac{m^2}{s})^n$. In other words, the soft regime cannot give rise to
a single logarithm at the one-loop level.

Consider an n-loop Feynman integral with loop momenta $l_i, 1 \le i \le n$.
We can decompose $l_i$ in terms of 'external momenta' $k_1$ and $k_2$,
\begin{equation}
l_i = \alpha_i k_1 + \beta_i k_2 + l_{i \perp}.
\end{equation} 
In the soft region, $|\alpha_i|, |\beta_i|, \frac{|l_{i \perp}^2|}{s} \ll 1$,
the Feynman integral does not give the next-to-leading logarithm,
$\log^{2n - 1} \frac{m^2}{s}$.
This is because, intuitively, each integration over the light
cone variables $\alpha$ and $\beta$ gives either a logarithm or a power
suppression, as exemplified in the previous paragraph. We now prove this
statement more rigorously.

We first note the following replacement for loop momenta corresponding to 
a soft line,
\begin{equation}
\frac{1}{l_i^2 - m_i^2 + i \epsilon} \rightarrow -i \pi
	\delta(s\alpha_i \beta_i + l_{i\perp}^2 - m_i^2).
\label{delta-func}
\end{equation}

The rest of the propagators can be categorized into hard, collinear
and soft ones.
The hard propagators are irrelevant to the infrared sensitivity.
The remaining possibilities are collinear to $k_1$ or $k_2$, soft or
collinear to $p$. We discuss them in turn.

(1) Collinear to $k_1$ or $k_2$. A boson that is parallel to $k_1$ 
has a propagator that can be written as
\begin{eqnarray}
\frac{1}{(k_1+\sum_i u_i l_i)^2} &=&
	\frac{1}{(1+\sum_i u_i \alpha_i)(\sum_i u_i \beta_i) s/2} \nonumber \\
& \rightarrow & \frac{2}{s} \frac{1}{\sum_i u_i \beta_i}.
\end{eqnarray}
The fermionic propagator can be expressed as
\begin{eqnarray}
\frac{1}{(\slk_1 + \sum_i u_i \sll_i) - m} & = &
	\frac{(\slk_1 + \sum_i u_i \sll_i) - m}{(k_1 + \sum_i u_i l_i)^2 - m^2}
	\nonumber \\
& \rightarrow & \frac{\slk_1}{(\sum_i u_i \beta_i)s/2}
\end{eqnarray}

(2) Soft. A bosonic propagator
\begin{equation}
\frac{1}{(\sum_i u_i l_i)^2} = \frac{1}{(\sum_i u_i \alpha_i)
	(\sum_i u_i \beta_i)s + (\sum_i u_i l_{i\perp})^2}
\end{equation}
Using $\delta(s \alpha_i \beta_i + l_{i\perp}^2 - m_i^2)$, and
$2\sqrt{(\alpha_i \beta_j)(\alpha_j \beta_i)} \le \alpha_i \beta_j + \alpha_j
\beta_i$, it is easy to show
\begin{equation}
|(\sum_i u_i \alpha_i)(\sum_i u_i \beta_i)s| > (\sum_i u_i l_{i\perp})^2.
\end{equation}
Hence, the expansion below,
\begin{equation}
\frac{1}{(\sum_i u_i \alpha_i)(\sum_i u_i \beta_i)s} \left [ 1 -
\frac{(\sum_i u_i l_{i\perp})^2}{(\sum_i u_i \alpha_i)(\sum_i u_i \beta_i)s}
	- \cdots \right ],
\label{soft}
\end{equation}
converges.
The fermionic propagator has a term proportional to $l_i^{\mu}$, in addition
to the expansion above.

(3) Collinear to $p$. Again, the bosonic propagator can be expanded as
\begin{eqnarray}
\frac{1}{(p + \sum_i u_i l_i)^2 - m_i^2} & \rightarrow &
	\frac{1}{\sum_i u_i (\alpha_i + \beta_i) |t|/2 +
	\sum_i u_i (p_{\perp} \cdot l_{i\perp})} \nonumber \\
& = & \frac{1}{\sum_i u_i (\alpha_i + \beta_i) |t|/2} \left [ 1 -
	\frac{\sum_i u_i (p_{\perp} \cdot l_{i\perp})}
		{\sum_i u_i (\alpha_i + \beta_i) |t|/2} - \cdots \right ]
\label{col-p}
\end{eqnarray}
The fermionic propagator may give rise to an additional $l_i^{mu}$ in
the numerator.

First, we only include the first term in the series in
eq.(\ref{soft}, \ref{col-p}) and apply the following trick
\begin{equation}
\frac{1}{\alpha} \frac{1}{\beta} \frac{1}{\alpha + \beta} = \left(
	\frac{1}{\alpha} - \frac{1}{\alpha + \beta} \right) \frac{1}{\beta^2}.
\end{equation}
This trick reduces the number of different combinations of $\alpha$'s and
$\beta$'s while splitting one term into two.

Therefore, when the propagators in case (1) and 
only the first terms in the expansions of the soft
and collinear-to-$p$ propagators are included, the integral can be reduced
to a finite sum of the type
\begin{equation}
\int_{\mathrm{soft \ regime}} \prod_{i}^{2n} d \sigma_i \frac{1}{\sigma^{m_i}},
\end{equation}
where $\sigma$'s represent $\alpha$'s, $\beta$'s or the combinations thereof.
It is evident that an integration
over each $\sigma$ gives either a logarithm or a power suppression $m^2/s$.
No next-to-leading logarithm can arise in the soft regime.

We now include the whole series in eq.(\ref{soft}, \ref{col-p}), as well as
their fermionic counterpart whenever appropriate.
The denominators are polynomials of $\alpha_i$ and
$\beta_i$,
\begin{equation}
\frac{1}{\sum_i u_i \alpha_i}, \frac{1}{\sum_i u_i \beta_i},
\frac{1}{\sum_i u_i (\alpha_i + \beta_i)}
\end{equation}
The numerators consists of terms proportional to $p^{\mu}$ and
$l_i^{\nu}$. The Feynman integral is thus of the following form
\begin{equation}
\int \prod d  \alpha_i d \beta_i d^2 l_{i\perp}
	\frac{\bar{f_1}(\alpha_i, \beta_i, p_{\perp} \cdot l_{i\perp},
	  l_{i\perp} \cdot l_{j\perp}) \prod l_{j\perp}^{\mu_j}}
	{\bar{f_2}(\alpha_i, \beta_i)},
\end{equation}
where $\bar{f_1}$ and $\bar{f_2}$ are both polynomial functions of their
arguments and the spinor structure is not interesting.
Since $p_{\perp}$ is the only vector in the integrand which is not the
integration variable, the integral of interest can further be reduced to
\begin{equation}
\int \prod d  \alpha_i d \beta_i d^2 l_{i\perp}
	\frac{\bar{f_1}(\alpha_i, \beta_i, p_{\perp} \cdot l_{i\perp},
	  l_{i\perp} \cdot l_{j\perp})}
	{\bar{f_2}(\alpha_i, \beta_i)},
\end{equation}
where we have further left out the tensorial structure of $p_{\perp}^{\mu}$
and $g^{\mu \nu}$.

By noting
\begin{eqnarray}
l_{i\perp} \cdot l_{j\perp} &=& \sqrt{l_{i\perp}^2} \sqrt{l_{j\perp}^2}
	\cos(\theta_i - \theta_j) \nonumber \\
p_{\perp} \cdot l_{i\perp} &= & \sqrt{p_{\perp}^2} \sqrt{l_{i\perp}^2}
	\cos \theta_i
\end{eqnarray}
where the angles $\theta_i$ are relative to the vector $\vec{p_{\perp}}$,
the integral can be cast into
\begin{eqnarray}
& \int \prod d  \alpha_i d \beta_i d l_{i\perp}^2
	\frac{f_1(\alpha_i, \beta_i, p_{\perp} \cdot l_{i\perp},
	  l_{i\perp} \cdot l_{j\perp})}
	{f_2(\alpha_i, \beta_i)} \int \prod d \theta_i f_3(\theta).
	\nonumber \\
\rightarrow & \int \prod d  \alpha_i d \beta_i \sum
	\frac{P(\sqrt{\alpha_i}, \sqrt{\beta_i},
	  \sqrt{\sum(u_i\alpha_i + v_j\beta_j)})}{Q(\alpha_i, \beta_i)}
\label{poly-sqrt}
\end{eqnarray}
where $\bar{P}$ and $\bar{Q}$ are polynomial functions of their arguments.
We have used the $\delta$ function (eq.(\ref{delta-func}))
to perform the integration over $l_{i\perp}^2$ and
implicitly included the resulting $\Theta$ functions in the
'polynomials' $\bar{P}$.
The summation in the second line is due to the expansions in the soft and
collinear-to-$p$ propagators. It is a convergent series.

Now we examine the arguments of the polynomial $\bar{P}$ in 
eq.(\ref{poly-sqrt}).
The $\sqrt{\sum(u_i\alpha_i + v_j\beta_j)}$ represents various combinations
of $\alpha_i$ and $\beta_j$ that may appear. For such a combination, we split
the integration region
\begin{equation}
\sqrt{\alpha_i + \beta_j} = \left \{
	\begin{array}{ll}
	\sqrt{\alpha_i} \left( 1 + \frac{1}{2} \frac{\beta_j}{\alpha_i} +
		\cdots \right), & \alpha_i > \beta_j \\
	\sqrt{\beta_j} \left( 1 + \frac{1}{2} \frac{\alpha_i}{\beta_j} +
		\cdots \right), & \alpha_i < \beta_j
	\end{array}
\right.
\end{equation}
After such maniputions, we obtain a convergent series
\begin{equation}
\int \prod d  \alpha_i d \beta_i \sum
	\frac{P(\sqrt{\alpha_i}, \sqrt{\beta_i})}{Q(\alpha_i, \beta_i)}
\end{equation}
with P and Q polynomial functions of $\sqrt{\alpha_i}$ and
$\sqrt{\beta_i}$. In order to obtain the
next-to-leading logarithm, $2n - 1$ of the integrations have to give
logarithm while the last one gives a constant of order 1.
This is impossible for the above integral in the soft region.
It follows that at least one of the loop momenta has to be taken out of
soft region to correctly reproduce the next to leading logarithmic
behavior in the Feynman gauge.

\subsection{A power counting technique}

In order to identify the regime contributing to the next-to-leading logarithmic
order at two-loop level,
we consider inserting a gluon into the one-loop box diagram.
From the previous subsection, we know the inserted gluon has to be collinear.
Therefore, we will consider in turn the three topologies with an additional
collinear gluon inserted to each of them.

Throughout the following subsections,
we always denote the soft loop momentum by $l$ and that of the collinear
gluon by $k$.

We take two momenta $p$ and $\bar{p}$ as the basis to decompose the
momentum $k$ of the collinear gluon,
\begin{equation}
k = \alpha p + \beta \bar{p} + k_{\perp}.
\end{equation}
The generic momenta $p$ and
$\bar{p}$ can be any two of the external momenta $k_1$, $k_2$,
$p_1$ and $p_2$.

In the so-called collinear region,
without loss of generality, we assume $k$ parallel to $p$ such
that 
\begin{equation}
|\alpha| \sim 1, |\beta| \sim \sqrt{\frac{k_{\perp}^2}{s}} \ll 1.
\end{equation}

In general, all the propagators in a Feynman diagram can be
characterized as hard (off-shell), soft or collinear (to a
certain direction). In order to get the next-to-leading
logarithm at two-loop level, which is a double logarithm multiplied with
a single logarithm, a Feynman diagram has to contain at least
four collinear and one soft propagators. Specifically,
the double logarithmic form factor arises from
a soft virtual particle 'interacting' with two (nearly) on-shell
particles that are flying apart along two different directions.
On the other hand, the interaction between two collinearly
flying virtual particles gives rise to the single logarithmic form factor.

As a result, whenever we have less than four collinear propagators,
we can immediately conclude that the Feynman diagram does not
contribute at the next-to-leading logarithmic level.
When we have exactly four collinear and a soft propagator, we only
keep terms proportional to $\alpha$ in the numerator. 
The $\beta$ term can be dropped because it cancels a $\beta$
in the denominator and effectively 'removes' a collinear propagator.
An example of such a collinear propagator is
\begin{equation}
\frac{1}{(p + k)^2 - m^2} = \frac{1}
  {(1 + \alpha) \beta (2 p \cdot \bar{p}) + k_{\perp}^2 + p^2 + 
	\bar{p}^2 - m^2}.
\end{equation}
And when we carry out the integration over $k$, we can pick up
a pole from an 'on-shell' propagator as the above one. It follows that
$\beta s \sim k_{\perp}^2$ and $k_{\perp}^2$ should also be dropped.
If we get more than four collinear propagators, we will
keep all the terms. However, in general the
terms proportional to $\alpha$ are suppressed by $m^2$ and thus
leave $\beta$ and $k_{\perp}^2$ the leading terms.

\subsection{Vertex functions}

When inserting a gluon into a box diagram, we will obtain
one-loop vertex subdiagrams inside four of the resulting two-loop diagrams.
Two of the vertex corrections each have two legs (nearly)
on-shell and the other soft. There is no large scale of order $s$, other
than that from the UV cutoff, in such a sub-diagram. Therefore,
these two vertex functions contain no infrared logarithms.

The other two vertex function subdiagrams have two (nearly) on-shell
and one off-shell legs each. They do contribute single logarithms
in all the three topologies.
Such vertex corrections
are shown in Fig.\ref{nll_a}(g, h), \ref{nll_b}(g, h), \ref{nll_c}(e, f)
and \ref{nll_cu}(e, f). Self energy corrections are understood and not
explicitly drawn.

Hereafter, we will omit the diagrams (and regions of diagram) that give
rise to large logarithms only of ultraviolet origin, until we are ready
to run the relevant parameters using the RGE.

\subsection{Contributing diagrams in topology A}

\begin{figure}
\centerline{\epsfig{file=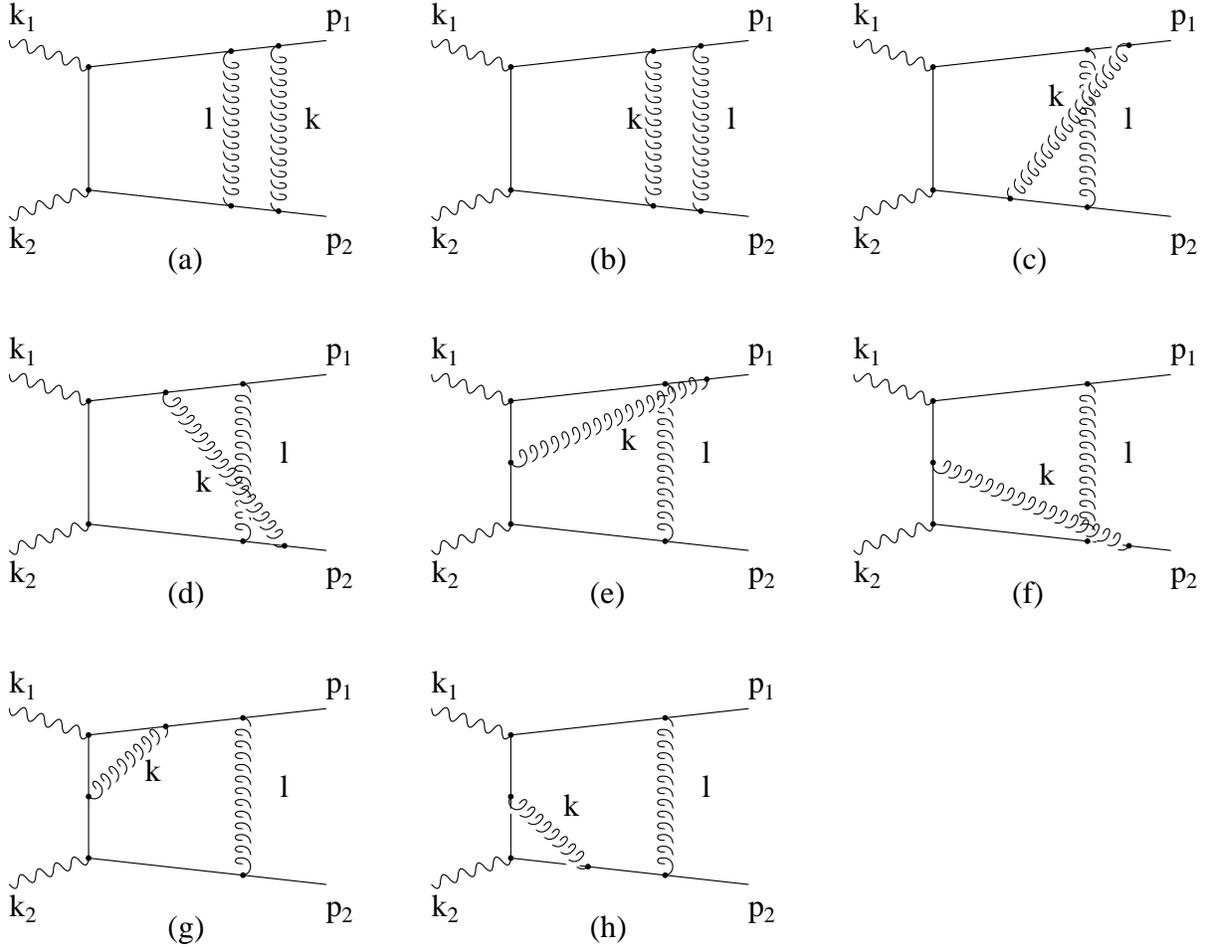, scale=0.85}}
\caption{\label{nll_a}
Diagrams relevant to the next-to-leading logarithms in Topology A.
The diagrams dependent only on $C_F$ are shown here.}
\end{figure}

We show some of the diagrams relevant to the next-to-leading logarithms 
of Topology A in Fig.\ref{nll_a}.
The shorter fermion lines in the box subdiagrams represent the off-shell
propagators, which characterize the Topology A.
Note Fig.\ref{nll_a}(a) and (b) represent different regions of the same
diagram, where the soft gluons are labeled by $l$ and the collinear
gluons by $k$. The same comments apply to Fig.\ref{nll_a}(c) and (d).
(We will follow the conventions that the characteristic
off-shell propagator is denoted by the shorter line, $l$ for the soft
momentum and $k$ for collinear momentum
throughout the rest of this article.)

\begin{figure}
\centerline{\epsfig{file=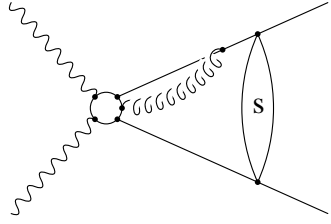, scale=0.85}}
\caption{\label{topa_jets}
Jet diagram for topology A}
\end{figure}

The reduced diagram for the first six diagrams, Fig.\ref{nll_a}(a)
through (f), all consist of a hard vertex with four jets attached to it,
as shown in Fig.\ref{topa_jets}. The two jets eventually emerge as quark
and antiquark further interact with each other via a soft diagram (the
gluon with momentum $l$). One of them consists of the collinear gluon
and the quark or antiquark. In addition, these subset of diagrams are
gauge invariant, since the reciprocal subset (consisting of vertex
correction and self-energy subdiagrams) are gauge invariant. We can now
invoke results from a general power counting analysis of infrared
sensitive contributions (both soft and collinear) to a typical wide-angle
scattering process \cite{Akhoury:1978vq,Sterman:bi}. 
It was shown
there that the logarithmic configuration requires that jet lines are
attached to hard vertices by a single line, otherwise there is power
suppression. The analysis of \cite{Akhoury:1978vq,Sterman:bi} was made
in a physical
gauge, but it obviously holds for the gauge invariant set discussed above.
Hence, the sum of Fig.\ref{nll_a}(a) through (f) does not contain infrared
logarithms. 

The remaining diagrams are Fig.\ref{nll_a}(b) and (h), which contribute
to the next-to-leading logarithmic order.
These are just the diagrams proportional only to $C_F$ contributing to
on-shell Sudakov form factor to NLL accuracy which is extensively
discussed in \cite{Collins:bt}

\subsection{Contributing diagrams in topology B}

\begin{figure}
\centerline{\epsfig{file=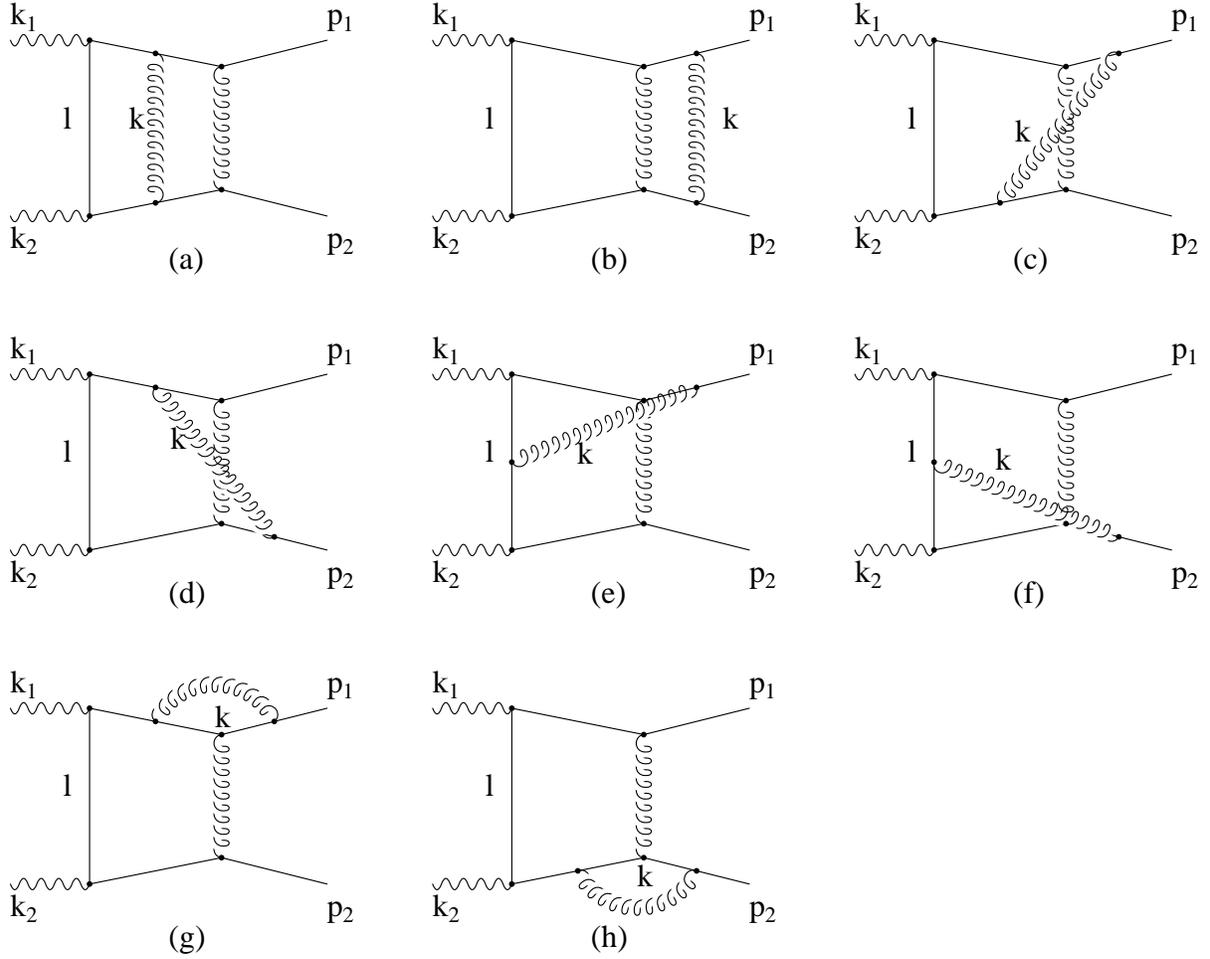, scale=0.85}}
\caption{\label{nll_b}
Diagrams relevant to the next-to-leading logarithms in Topology B.
The t-channel diagrams dependent only on $C_F$ are shown here.}
\end{figure}

Now we turn to the diagrams in Fig.(\ref{nll_b}).
In Fig.(\ref{nll_b}e), the collinear gluon, labeled
by $k$, can be parallel to $p_1$, $k_1$ or $k_2$. We discuss them in turn.

i) Parallel to $p_1$ ($k \| p_1$). There are only three collinear
propagators left. This region can be excluded by power counting.

ii) Parallel to $k_1$ ($k \| k_1$). The soft fermion is the one
labeled by $l$, while the fermion labeled by $k + l$ is collinear to
$k_1$. And the fermion with momentum $k + p_1$ is off-shell. 
There are exactly four collinear and one soft propagator left. Hence we
only keep the component of $k$ that is parallel to $k_1$.
The numerator of the diagram here is,
\begin{eqnarray}
N & \propto & (\slk_1 + \slk + \sll) \slep(\slk_1) (\slk + \sll)
	\nonumber \\
& \propto & \slk_1 \slep(\slk_1) \slk_1 = 0,
\end{eqnarray}
which implies a vanishing contribution to the next-to-leading logarithmic
order from this region.

iii) The last possible region is $k \| k_2$, which vanishes due
to the similar reason as in ii).

Therefore, Fig.\ref{nll_b}(e) vanishes as well as Fig.\ref{nll_b}(f).
Diagrams Fig.\ref{nll_b}(a) - (d) can be shown to factorize. 
Take Fig.\ref{nll_b}(d) as an example, the numerator of the amplitude is
\begin{eqnarray}
N_{16d} & = & \bar{u}(p_1) \gamma^{\alpha} (\slk_1 + \sll + \slk) \gamma^{\mu}
  (\slk_1 + \sll) \slep(k_1) m \slep(k_2) (-\slk_2 + \sll) \gamma_{\alpha}
  (-\slp_2 + \slk) \gamma_{\mu} v(p_2)	\nonumber \\
& = & \left \{
  \begin{array}{ll}
    \bar{u}(p_1) \gamma^{\alpha} (\slk_1 + \sll) \slep(k_1) m \slep(k_2)
	(-\slk_2 + \sll) \gamma_{\alpha} \left[ -4 p_2 \cdot (k_1 + k) \right]
	v(p_2), & k \| k_1 \\
   \bar{u}(p_1) \gamma^{\alpha} \left[ 4 k_1 \cdot (-p_2 + k) \right]
	(\slk_1 + \sll)	\slep(k_1) m \slep(k_2)(-\slk_2 + \sll)	\gamma_{\alpha}
	v(p_2), & k \| p_2 
  \end{array}
\right. 
\end{eqnarray}
The factorization is evident now. Similar results hold for the other three
diagrams.
However, they do not contain any large single logarithms.

The remaining diagrams, Fig.\ref{nll_b}(g) and (h) are just the ones
included in the resummation discussed in Section 2.3 to this order.

\subsection{Contributing diagrams in topology C}

\begin{figure}
\centerline{\epsfig{file=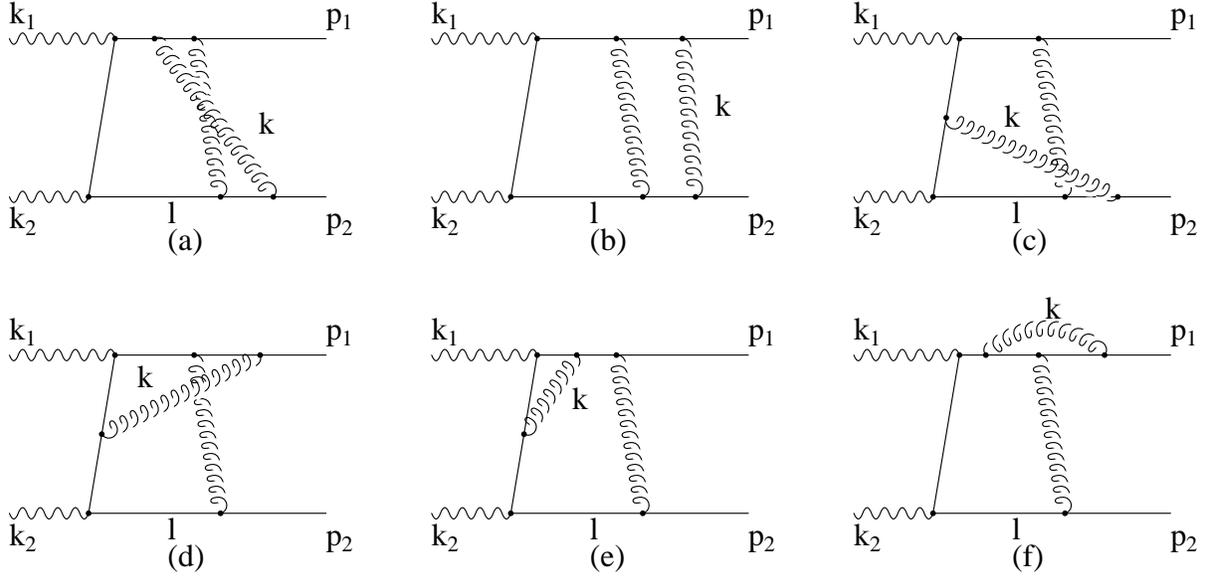, scale=0.85}}
\caption{\label{nll_c}
Diagrams relevant to the next-to-leading logarithms in Topology C.
The s-channel diagrams dependent only on $C_F$ are shown here.}
\end{figure}

In topology C,
the two-loop s-channel diagrams only proportional to $C_F$
are shown in Fig.(\ref{nll_c}).
The regions contributing to the next-to-leading logarithmic approximation
for Fig.(\ref{nll_c}a, b, c) are both the gluons being parallel to $p_2$.

Note the 'incoming' quark and the 'outgoing' gluon of the hard subprocess
$\gamma q \rightarrow qg$ are nearly on shell.
In addition, the gluon labeled by $k$ in the three diagrams are nearly
on-shell too.
We can expect the contributions only proportional to $C_F$ of
the three diagrams to cancel to the leading order, in
the same manner as discussed earlier in Section 3.

In order to show the cancellation, we decompose
\begin{equation}
k = \alpha_k k_2 + \beta_k p_2 + k_{\perp}.
\end{equation}
Each of the three amplitude takes on the form
\begin{eqnarray}
M_i &=& \bar{M_i}^{\mu} \frac{1}{- \slp_2 - \slk} \gamma_{\mu} v(p_2) \\
\nonumber
 & = & - \bar{M_i}^{\mu} \frac{\alpha_k \slk_2 + (1+\beta_k) \slp_2 +
\slk_{\perp}}{
	(p_2+k)^2} \gamma_{\mu} v(p_2),
\end{eqnarray}
with $M_i$, $i=1, 2, 3$, representing amplitudes of
Fig.(\ref{nll_c}a, b, c) respectively.
In this subsection, integrations over $l$ and $k$ should be understood in the
amplitudes. We consistently omit common numerical factors for simplicity.
To the single logarithmic approximation, $\alpha_k$ and $k_{\perp}$ terms in
the numerators can be neglected. Hence,
\begin{eqnarray}
M_i &=& - \bar{M_i}_{\mu} \frac{(1+\beta_k)p_2^{\mu}}{(p_2+k)^2} v(p_2) \\
\nonumber
 & = & - \frac{1+\beta_k}{\beta_k} \bar{M_i}_{\mu} k^{\mu} \frac{1}{(p_2+k)^2}
v(p_2) 
\end{eqnarray}
and we have put back the $\alpha_k k_2 + k_{\perp}$ in the second line.

The summation of $k^{\mu} \bar{M_i}_{\mu}$ in the three amplitudes
closely follows
the earlier proof using the Ward Identity (see section 3.1). We obtain
(for the piece proportional only to $C_F$)
\begin{eqnarray}
M_1 + M_2 +M_3 = & - \frac{1+\beta_k}{\beta_k} [ \bar{u}(p_1) \slp_1
	\gamma^{\lambda}
	 \frac{1}{\slk_1+\slk_2+\sll} \slep(k_1) \frac{1}{\slk_2+\sll}
	  - \bar{u}(p_1) \gamma^{\lambda} \frac{1}{\slk_1+\slk_2+\sll+\slk}
	  \slep(k_1) \frac{1}{\slk_2+\slk+\sll} ] \\ \nonumber
	& \cdot \slep(k_2)
	  \frac{1}{\sll-m} \gamma_{\lambda} v(p_2) \frac{1}{(p_2+k)^2}
	  \frac{1}{k^2} \frac{1}{(p_2+k+l)^2}.
\end{eqnarray}
The first term is suppressed by the quark mass, while the second one vanishes
in the single logarithmic approximation by simple power counting. Therefore,
the contributing diagram in Fig.{\ref{nll_c}} are d, e, and f.

\begin{figure}
\centerline{\epsfig{file=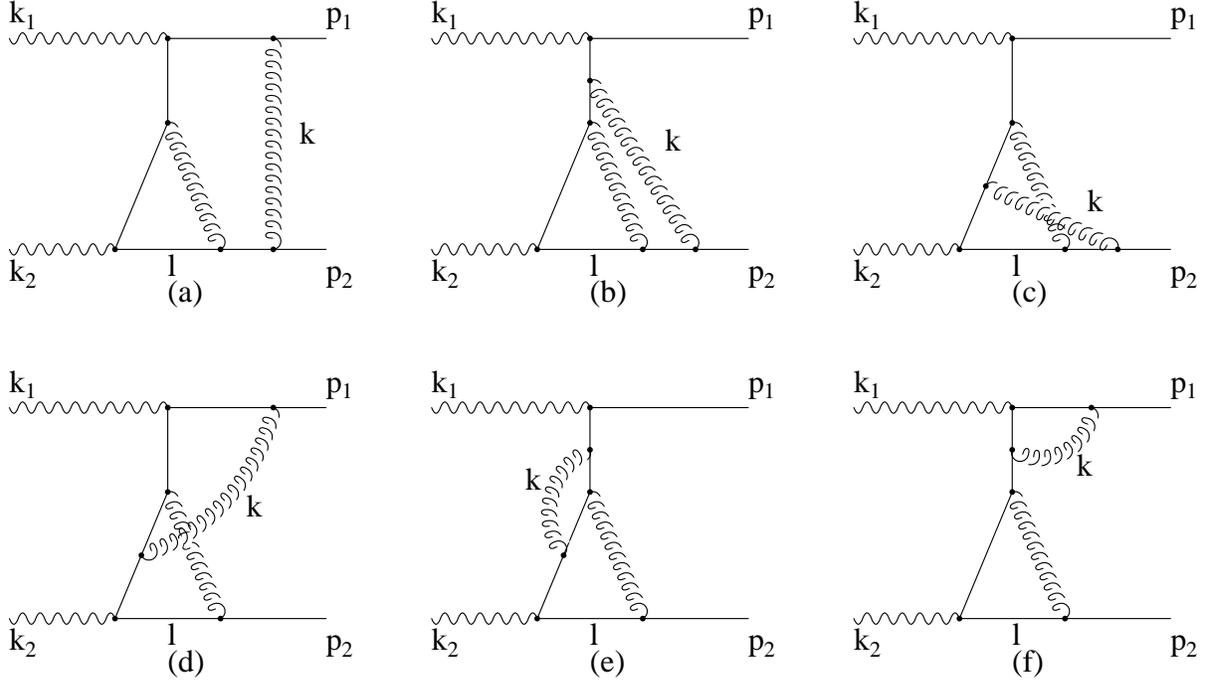, scale=0.85}}
\caption{\label{nll_cu}
Diagrams relevant to the next-to-leading logarithms in Topology C.
The u-channel diagrams dependent only on $C_F$ are shown here.}
\end{figure}

Similarly, the diagrams with the u-channel hard subprocess also include six
diagrams.
Noteworthy is that Fig.\ref{nll_cu}(a) can also be drawn as a vertex
correction to the s-channel diagrams, with the triangle subdiagram being
the vertex correction therein. However, we notice they represent different
regions. In Fig.\ref{nll_cu}(a), the quark labeled by $l$ is soft,
whereas in the s-channel diagram, the corresponding one is collinear to $k_2$.
A similar remark applies to Fig.\ref{nll_cu}(b). Here, it can be viewed
as a vertex correction to the u-channel diagram, which in turn represents
a different region.
As a summary, the two diagrams shown in Fig.\ref{nll_cu}(a, b) represent
the region that both the virtual gluons are parallel to $p_2$, while in their
counterparts, the two virtual gluons are parallel to $k_2$ and $p_2$
respectively.
Obviously, the sum of Fig.\ref{nll_cu}(a, b, c) vanishes to the next-to-leading
logarithmic
approximation in exactly the same way Fig.\ref{nll_c}(a, b, c) do.

\begin{figure}
\centerline{\epsfig{file=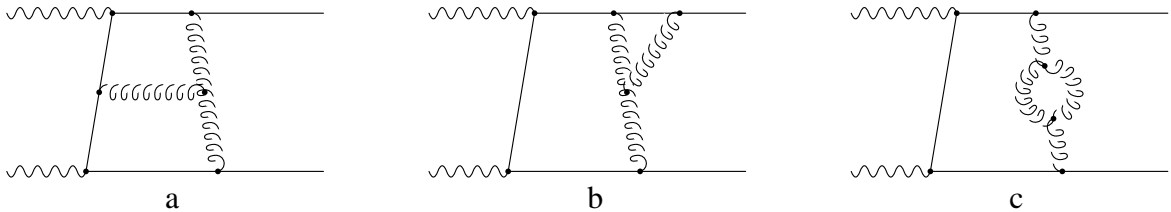, scale=0.85}}
\caption{\label{topc_nab}
Some of the diagrams proportional to $C_A$ in topology C.
}
\end{figure}

Therefore, the contributing diagrams of the Topology C at the two-loop level
include Fig.\ref{nll_c}(d, e, f) and Fig.\ref{nll_cu}(d, e, f) only. We
note the diagrams contributing to the ``hard'' off-shell Sudakov form
factor in this topology are mostly non-abelian which are of Sudakov
type. Some of these are shown in Fig.\ref{topc_nab}. This completes our
argument justifying that the only source of logarithm at this order are
of the ``Sudakov'' type.

\subsection{Extension to higher loops}

In this section, we argue without detailed proof that the above results 
hold to
the full NLL accuracy. We consider only topologies B and C, since
topology A, which corresponds to the on-shell Sudakov form factor, has
been discussed in detail in \cite{Collins:bt}.

For topologies B and C, consider the insertions of a gluon into the 
bare diagram. There are two cases: \\
i) The gluon momentum is in the soft region. The analysis in Section 3
can be applied and it can be factored out.\\
ii) If the gluon momentum is in collinear region, the analysis in the
previous subsections apply and again, factorization results.

All subsequent insertions of gluons in case (ii) must be restricted to
the 
soft region. Therefore, for these the analysis of section 3 apply. In
case (i), we keep inserting gluons and apply the same analysis until we
encounter a collinear gluon. Then case (ii) applies.

Hence we can conclude that to NLL accuracy, the relevant diagrams are
all of the Sudakov type.

\section{Summary and Discussions }

In this paper we have studied the resummation up to the next to leading 
logarithmic level
of the QCD radiative corrections to $\bar{b}b$ production by photon
photon collisions.
Apart from the phenomenological applications, this problem has inherent 
interest in providing a theoretical laboratory to study QCD effects. We 
showed that to
the accuracy considered all logarithms are of the Sudakov type. On a
diagram by diagram
basis other types of diagrams do give rise to next to leading logarithms
but they
cancel amongst each other by the dipole mechanism. We explicitly showed
how the dipole mechanism
works to 3 loops and outlined an all orders generalization. At the NLL 
level we found that in the
Feynman gauge the only new logarithms are of collinear origin. The final
expression for the resummed
amplitude is given in section 2, eq.(\ref{ampNLL}) together with 
eq.(\ref{rep1},\ref{finalB},\ref{finalC}). 

\section{Acknowledgments} 

We would like to thank G. Sterman for discussions.
This work was supported by the US Department of Energy.


\begin{thebibliography}{9}

%\cite{Gunion:1989we}
\bibitem{Gunion:1989we} J.F. Gunion, H.E. Haber, G. Kane, and S. Dawson,
The Higgs Hunter's Guide (Addison-Wesley, Reading, MA, 1990).

%\cite{Gunion:1992ce}
\bibitem{Gunion:1992ce}
J.~F.~Gunion and H.~E.~Haber,
%``Higgs boson production in the photon-photon collider mode of a high-energy e+ e- linear collider,''
Phys.\ Rev.\ D {\bf 48}, 5109 (1993).
%%CITATION = PHRVA,D48,5109;%%

%\cite{Hagiwara:fs}
\bibitem{Hagiwara:fs}
K.~Hagiwara {\it et al.}  [Particle Data Group Collaboration],
%``Review Of Particle Physics,''
Phys.\ Rev.\ D {\bf 66}, 010001 (2002).
%%CITATION = PHRVA,D66,010001;%%

\bibitem{QCDEW} 
K.~Melnikov and O.~Yakovlev,
%``Higgs $\to$ two photon decay: QCD radiative correction,''
Phys.\ Lett.\  {\bf B312}, 179 (1993) [hep-ph/9302281].\\
A.~Djouadi, M.~Spira, J.~J.~van der Bij and P.~M.~Zerwas,
%``QCD corrections to gamma gamma decays of Higgs particles in the intermediate mass range,''
Phys.\ Lett.\  {\bf B257}, 187 (1991).\\
J.~G.~Korner, K.~Melnikov and O.~I.~Yakovlev,
%``Two loop O(G(f)M(H**2)) radiative corrections to the Higgs decay width H $\to$ gamma gamma for large Higgs boson masses,''
Phys.\ Rev.\  {\bf D53}, 3737 (1996) hep-ph/9508334]. \\
K.~Melnikov, M.~Spira and O.~Yakovlev,
%``Threshold effects in two photon decays of Higgs particles,''
Z.\ Phys.\  {\bf C64}, 401 (1994)


%\cite{Ginzburg:1981vm}
\bibitem{Ginzburg:1981vm}
I.~F.~Ginzburg, G.~L.~Kotkin, V.~G.~Serbo and V.~I.~Telnov,
%``Colliding Gamma E And Gamma Gamma Beams Based On The Single Pass Accelerators (Of Vlepp Type),''
Nucl.\ Instrum.\ Meth.\  {\bf 205}, 47 (1983).
%%CITATION = NUIMA,205,47;%%

%\cite{Telnov:hc}
\bibitem{Telnov:hc}
V.~Telnov,
%``Principles Of Photon Colliders,''
Nucl.\ Instrum.\ Meth.\ A {\bf 355}, 3 (1995).
%%CITATION = NUIMA,A355,3;%%

%\cite{Fadin:1997sn}
\bibitem{Fadin:1997sn}
V.~S.~Fadin, V.~A.~Khoze and A.~D.~Martin,
%``Higgs studies in polarized gamma gamma collisions,''
Phys.\ Rev.\ D {\bf 56}, 484 (1997)
[arXiv:hep-ph/9703402].
%%CITATION = HEP-PH 9703402;%%

%\cite{Borden:1994fv}
\bibitem{Borden:1994fv}
D.~L.~Borden, V.~A.~Khoze, W.~J.~Stirling and J.~Ohnemus,
%``Three jet final states and measuring the gamma gamma width of the Higgs at a photon linear collider,''
Phys.\ Rev.\ D {\bf 50}, 4499 (1994)
[arXiv:hep-ph/9405401].
%%CITATION = HEP-PH 9405401;%%
%%%%%%%%%%%%%%%%

%\cite{Jikia:1996bi}
\bibitem{Jikia:1996bi}
G.~Jikia and A.~Tkabladze,
%``QCD Corrections to Heavy Quark Pair Production in Polarized $\gamma\gamma$ Collisions and the Intermediate Mass Higgs Signal,''
Phys.\ Rev.\ D {\bf 54}, 2030 (1996)
[arXiv:hep-ph/9601384].
%%CITATION = HEP-PH 9601384;%%

%\cite{Melles:1998gu}
\bibitem{Melles:1998gu}
M.~Melles and W.~J.~Stirling,
%``All-orders resummation of leading logarithmic contributions to heavy  quark production in polarized gamma gamma collisions,''
Phys.\ Rev.\ D {\bf 59}, 094009 (1999)
[arXiv:hep-ph/9807332].
%%CITATION = HEP-PH 9807332;%%

%\cite{Sudakov:1954sw}
\bibitem{Sudakov:1954sw}
V.~V.~Sudakov,
%``Vertex Parts At Very High-Energies In Quantum Electrodynamics,''
Sov.\ Phys.\ JETP {\bf 3}, 65 (1956)
[Zh.\ Eksp.\ Teor.\ Fiz.\  {\bf 30}, 87 (1956)].
%%CITATION = SPHJA,3,65;%%

%\cite{Gorshkov:ht}
\bibitem{Gorshkov:ht}
V.~G.~Gorshkov, V.~N.~Gribov, L.~N.~Lipatov and G.~V.~Frolov,
%``Doubly Logarithmic Asymptotic Behavior In Quantum Electrodynamics,''
Sov.\ J.\ Nucl.\ Phys.\  {\bf 6}, 95 (1968)
[Yad.\ Fiz.\  {\bf 6}, 129 (1967)].
%%CITATION = SJNCA,6,95;%%

%\cite{Carazzone:hj}
\bibitem{Carazzone:hj}
J.~Carazzone, E.~C.~Poggio and H.~R.~Quinn,
%``Asymptotic Behavior Of Form-Factor In Nonabelian Gauge Theories,''
Phys.\ Lett.\ B {\bf 57}, 161 (1975).
%%CITATION = PHLTA,B57,161;%%

%\cite{Collins:bt}
\bibitem{Collins:bt}
J.~C.~Collins,
%``Sudakov Form-Factors,''
Adv.\ Ser.\ Direct.\ High Energy Phys.\  {\bf 5}, 573 (1989).
%%CITATION = 00319,5,573;%%

%\cite{Smilga:uj}
\bibitem{Smilga:uj}
A.~V.~Smilga,
%``Next-To-Leading Logarithms In The High-Energy Asymptotics Of The Quark Form-Factor And The Jet Cross-Section,''
Nucl.\ Phys.\ B {\bf 161}, 449 (1979).
%%CITATION = NUPHA,B161,449;%%

%\cite{Akhoury:2001mz}
\bibitem{Akhoury:2001mz}
R.~Akhoury, H.~Wang and O.~I.~Yakovlev,
%``On the resummation of large QCD logarithms in H $\to$ gamma gamma,''
Phys.\ Rev.\ D {\bf 64}, 113008 (2001)
[arXiv:hep-ph/0102105].
%%CITATION = HEP-PH 0102105;%%

\bibitem{Sterman:Book} G. Sterman, An Introduction to Quantum Field
	Theory, Cambridge University Press, 1993.

%\cite{Libby:nr}
\bibitem{Libby:nr}
S.~B.~Libby and G.~Sterman,
%``Cancellation Of Infrared Divergences In Massive Quark Potential Scattering,''
Phys.\ Rev.\ D {\bf 19}, 2468 (1979).\\
%%CITATION = PHRVA,D19,2468;%%
V.~Ganapathi and G.~Sterman,
%``Infrared Divergences In Quark Potential Scattering,''
Phys.\ Rev.\ D {\bf 23}, 2408 (1981).
%%CITATION = PHRVA,D23,2408;%%

%\cite{Akhoury:1978vq}
\bibitem{Akhoury:1978vq}
R.~Akhoury,
%``Mass Divergences Of Wide Angle Scattering Amplitudes,''
Phys.\ Rev.\ D {\bf 19}, 1250 (1979).
%%CITATION = PHRVA,D19,1250;%%

%\cite{Sterman:bi}
\bibitem{Sterman:bi}
G.~Sterman,
%``Mass Divergences In Annihilation Processes.  1.  Origin And Nature Of Divergences In Cut Vacuum Polarization Diagrams,''
Phys.\ Rev.\ D {\bf 17}, 2773 (1978).
%%CITATION = PHRVA,D17,2773;%%

\end{thebibliography}
\end{document}